\documentclass[aps,pra,10pt,superscriptaddress,twocolumn,longbibliography]{revtex4-1}

\usepackage{color,graphicx,amsmath}

\usepackage[pdfpagelabels,pdftex,bookmarks,breaklinks]{hyperref}
\definecolor{darkpurple}{RGB}{100,0,120}
\definecolor{darkgreen}{RGB}{0,150,0}
\definecolor{darkblue}{RGB}{0,0,130}
\hypersetup{colorlinks=true, linkcolor=darkpurple, citecolor=darkgreen, filecolor=red, urlcolor=darkblue}
\hypersetup{pdftitle={Adiabatic Quantum Transistors}}
\hypersetup{pdfauthor={Dave Bacon, Steven T. Flammia, Gregory M. Crosswhite}}

\begin{document}

\title{Adiabatic Quantum Transistors} 
\author{Dave Bacon}
\altaffiliation[Current affiliation: ]{Google Inc.}
\affiliation{Department of Computer Science and Engineering, University of Washington, Seattle, WA 98195 USA}
\affiliation{Department of Physics, University of Washington, Seattle, WA 98195 USA}
\author{Steven T. Flammia}
\affiliation{School of Physics, University of Sydney, Sydney NSW 2006 Australia}
\affiliation{Department of Computer Science and Engineering, University of Washington, Seattle, WA 98195 USA}
\author{Gregory M. Crosswhite}
\altaffiliation[Current affiliation: ]{Department of Physics, University of Queensland, Brisbane, QLD 4072 Australia}
\affiliation{Department of Physics, University of Washington, Seattle, WA 98195 USA}

\begin{abstract}
We describe a many-body quantum system which can be made to quantum compute by the adiabatic application of a large applied field to the system. Prior to the application of the field quantum information is localized on one boundary of the device, and after the application of the field this information has propagated to the other side of the device with a quantum circuit applied to the information. The applied circuit depends on the many-body Hamiltonian of the material, and the computation takes place in a degenerate ground space with symmetry-protected topological order. Such \textit{adiabatic quantum transistors} are universal adiabatic quantum computing devices which have the added benefit of being modular. Here we describe this model, provide arguments for why it is an efficient model of quantum computing, and examine these many-body systems in the presence of a noisy environment.
\end{abstract}
\maketitle

\section{Introduction}

The invention of the transistor~\cite{Shockley:48a} was a watershed moment in the history of computing: it provided a logic element that was naturally robust to noise and error. Quantum computers offer the potential to exponentially speed up some computational problems (notably factoring~\cite{Shor:94a}), but have not been built in large part because quantum information is notoriously fragile and quickly becomes classical information in the presence of noise. In theory, the quantum threshold theorem~\cite{Aharonov:97a, Knill:98a, Knill:98b} asserts that these difficulties can be circumvented, but in practice the requirements of this theorem are daunting. Here we outline a novel method for building a fault-tolerant quantum computer that much more closely mimics the classical transistor. In particular, we show how a suitably engineered material can quantum compute by the simple application of an external field to the sample. Applying the field adiabatically causes quantum information to spatially propagate across the device at the same time that a quantum computation (quantum circuit) is enacted on the quantum information, and this in turn allows us to design clocked quantum computing architectures similar in control requirements to modern classical computers. While we will not be able to rigorously show that our adiabatic quantum transistors are fault-tolerant devices like classical transistors, we present analytical and numerical arguments as to why these transistors could be tolerant to errors and thus a true quantum analog of the classical transistor. 

The standard operating model for a quantum computer is called the \emph{quantum circuit model}~\cite{Deutsch:89a}. In this model, one begins with a system of initialized two-level quantum systems (qubits), applies a temporal sequence of one and two-qubit \emph{gates} enacting a circuit, and finally performs a measurement (readout) of the qubits. In contrast with modern classical computers where information propagates spatially---advancing roughly one step across the computer chip at each rise and fall of a clock voltage---most proposed implementations of quantum computers fix the information spatially and bring the computational operations to the data. Many researchers have noted this difference, and a variety of quantum computing (QC) models were subsequently developed where quantum information propagates spatially during a quantum computation; examples include spin-wave models where quantum information moves ballistically down a quantum wire~\cite{Bose:03a}, linear-optics QC~\cite{Knill:01a}, one-way QC where simple sequential measurements push the quantum information across the system~\cite{Raussendorf:01a}, and some universal adiabatic QC models which create a superposition of computational states spread across the device~\cite{Aharonov:04a, Kempe:06a, Mizel:06a}. However none of these constructions yield architectures that mimic modern synchronous sequential computer chips. In these chips the rise or fall of a global voltage is the trigger that causes the information in the device to move spatially across the device. While the control requirements for the clock signal in synchronous sequential logic devices are by no means trivial, they do not require the precise control, measurement, and timing that makes quantum computers notoriously difficult to build. Here we introduce a new way to mimic these classical control requirements that allows us to propose synchronous sequential fault-tolerant QC architectures. The key to our construction is a novel type of quantum gadget that mimics the role classical transistors play in modern computers, and that we hence label an adiabatic quantum transistor. While the outline we give for an adiabatic quantum transistor is still very far from experimental realization, the novel quantum computational matter that such a device represents opens a new and potentially promising path for the construction of a large scale quantum computer.

An outline of the paper is as follows. In Sec.~\ref{sec:aqt} we introduce a model of an adiabatic quantum transistor. The speed at which we can operate a quantum transistor is related to the minimal energy gap between the ground state manifold and first excited state during the application of the applied field. In Sec.~\ref{sec:1d} we rigorously prove that if the adiabatic quantum transistor enacts a single-qubit quantum circuit with only identity gates, then the energy gap is such that the adiabatic quantum transistor can be enacted efficiently. In Sec.~\ref{sec:1dgates} we provide strong numerical evidence from matrix product state simulations that the same efficiency holds when the adiabatic quantum transistor enacts an arbitrary single-qubit quantum circuit. Next in Sec.~\ref{sec:2d} we provide (weaker) numerical evidence that the same results hold for our adiabatic quantum transistors for quantum circuits involving more than one qubit. Having given evidence that the adiabatic quantum transistor model is an efficient model of quantum computation in an ideal world in which the system is isolated from its environment, in Sec.~\ref{sec:ft} we turn to the question of whether the model can be made fault-tolerant. We give arguments that if quantum transistors are configured to execute fault-tolerant quantum circuits then the tolerance of these circuits to errors is conveyed to our model. While we cannot show that our model has a threshold theorem associated with it, we can give physical reasons for why the model will have a threshold. In Sec.~\ref{sec:gadgets} we discuss how the unrealistic four-body Hamiltonian we use in our constructions can be implemented using a Hamiltonian with only two-body interactions, via perturbation theory gadgets, and discuss the effect that these gadgets have on the arguments in the previous sections. In Sec.~\ref{sec:blocks} we present a variety of constructions for quantum adiabatic transistors that are likely to be useful, including systems that can be used to spatially route quantum information and to perform measurements. Finally in Sec.~\ref{sec:conc} we conclude and list important open questions for the adiabatic quantum transistor model.

\section{The Adiabatic Quantum Transistor Model}\label{sec:aqt}

\begin{figure}[t]
\begin{center}
\includegraphics[width=3.5in]{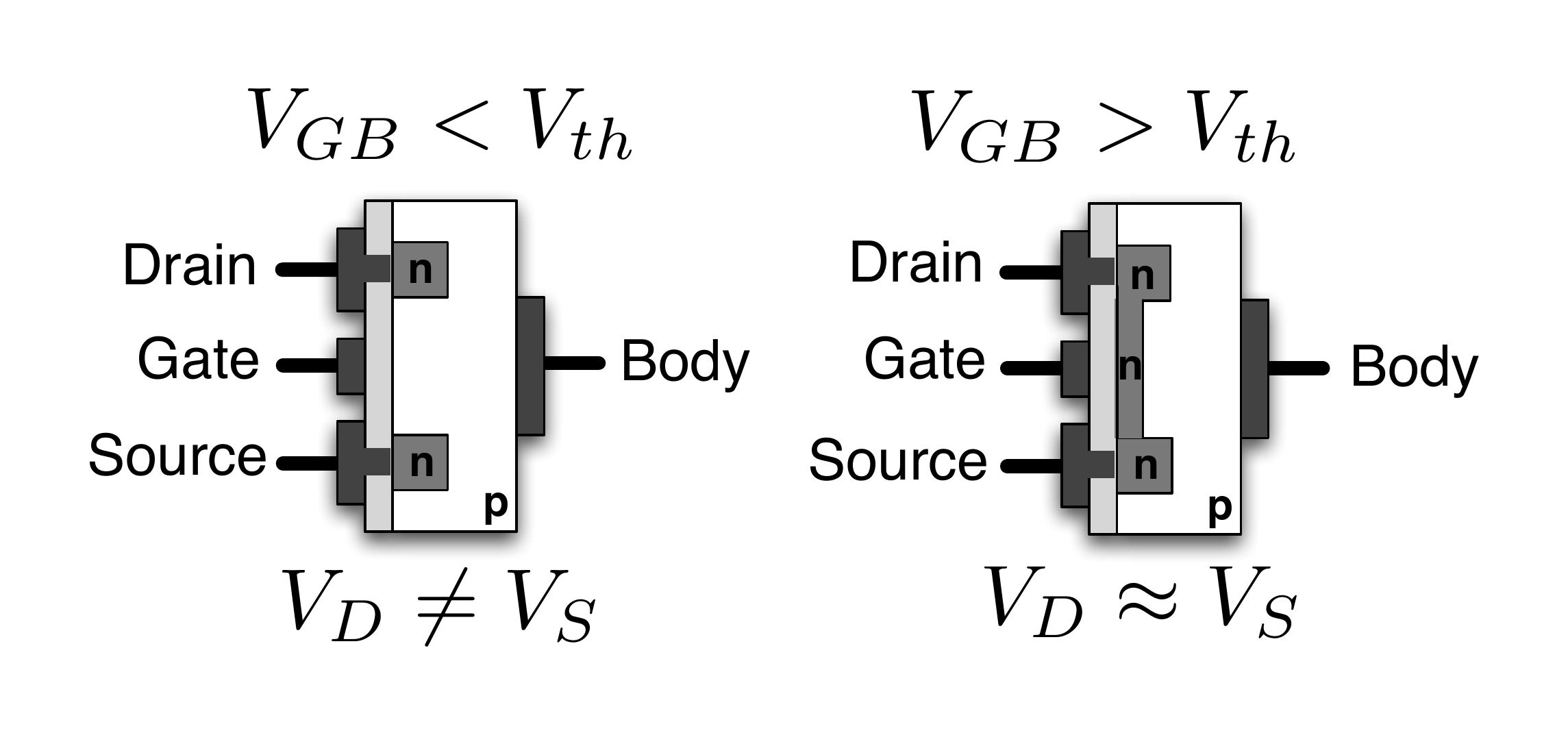}
\caption{Graphical depiction of the operation of a classical transistor. In a classical transistor such as the $n$-channel MOSFET shown, the application of a voltage at the gate input transforms the semiconductor in a manner such that a channel opens between drain and source. Note that this transition is switched by the application of an electric field. Traditionally we think of this action as a switch, but there is another possible interpretation. In particular, the application of the applied electric field has the effect of \textit{copying} the voltage at the source to the drain, which we can interpret as an \textit{identity} logical gate (a logical gate that takes in a single bit and outputs this bit unchanged) on the classical information encoded as a voltage. Quantum information cannot be copied~\cite{Wootters:82a}, so if we are to imagine a suitable analogy of the transistor in the quantum world, the quantum information must move from the source to drain. This is what our proposed adiabatic quantum transistors achieve.} 
\label{fig:transistor}
\end{center}
\end{figure}

\begin{figure}[t]
\begin{center}
\includegraphics[width=3.5in]{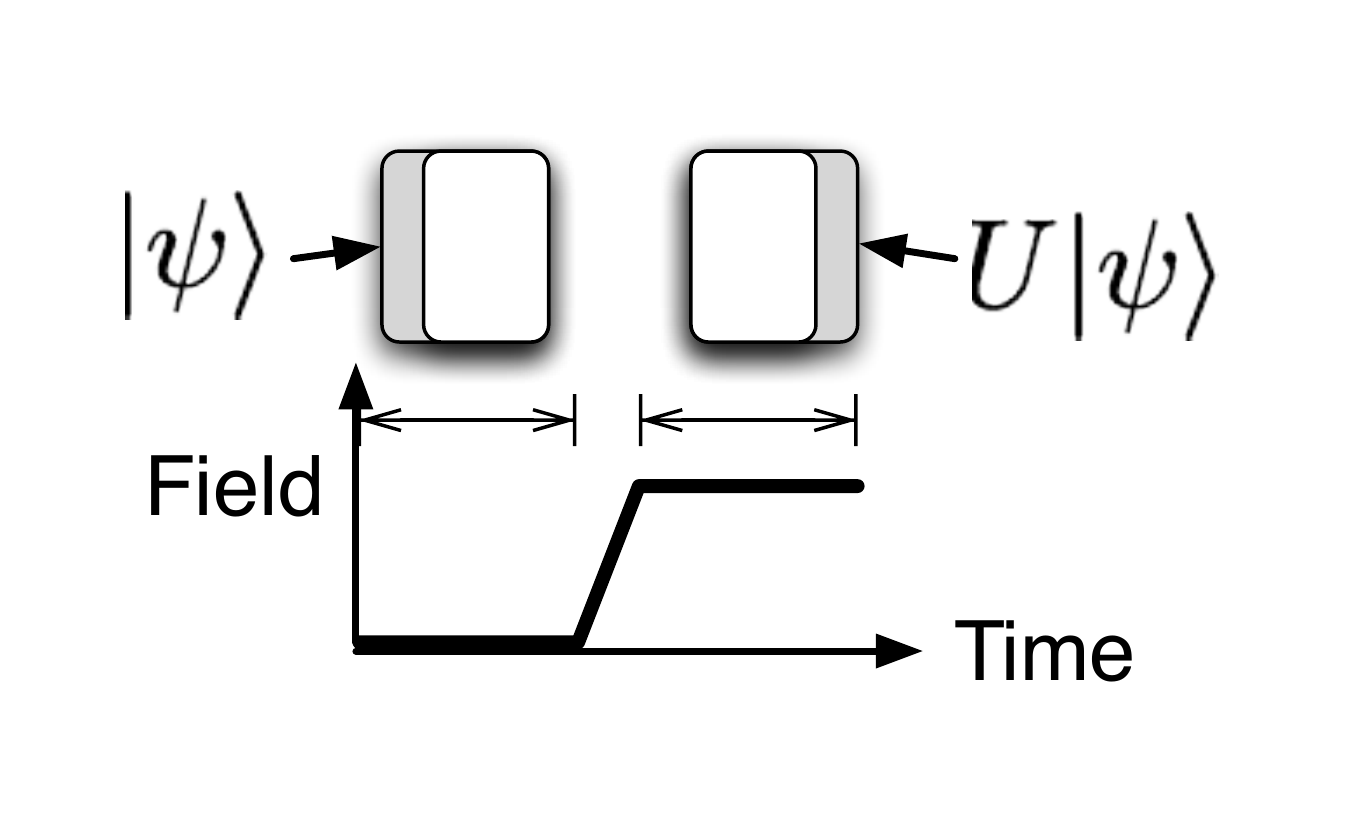}
\caption{Sketch of how an adiabatic quantum transistor behaves. With no applied field, the quantum information is on one side of the device (shaded grey, left) and after the adiabatic application of the field the quantum information has propagated to the other side of the device (shaded grey, right) with a unitary $U$ applied to the original information. The strength of the applied field is initially zero and then is ramped up to a non-zero value, during which time the quantum information is not localized to either end of the device.}
 \label{fig:aqt}
\end{center}
\end{figure}

\begin{figure}[ht]
\begin{center}
\includegraphics[width=3.5in]{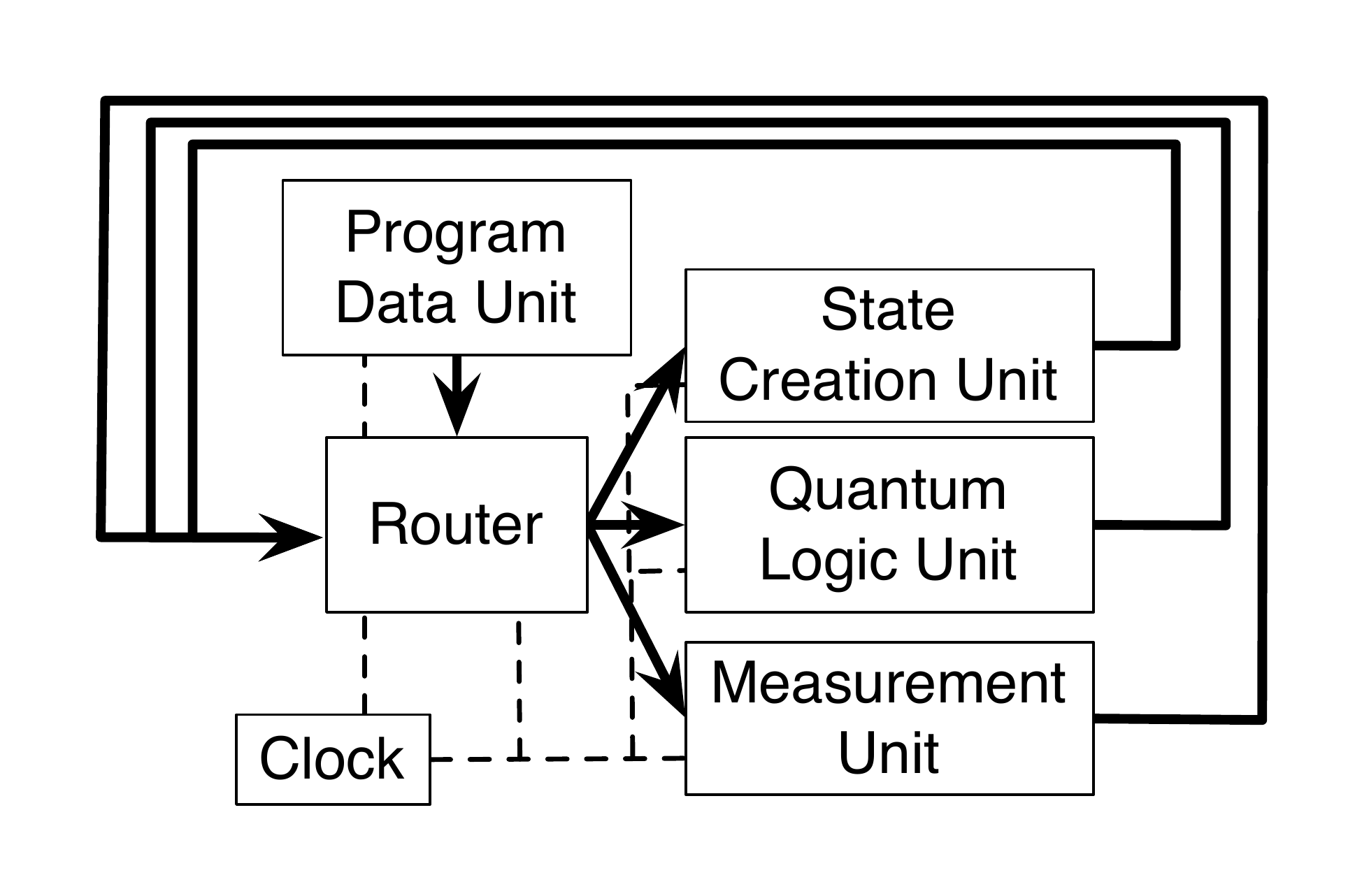}
\caption{Using quantum transistors one can design a synchronous architecture as shown. A router takes classical information about what circuit should be executed (the quantum program) and routes it to the appropriate action unit. These units are made up of a collection of quantum transistors that can be used to implement their given functions. This information is then directed to the router where another step in the circuit begins by reading the next program action.}
\label{fig:clock}
\end{center}
\end{figure}

Our proposed adiabatic quantum transistor is a device that operates in a manner similar to a classical logical element such as a MOSFET. See Fig.~\ref{fig:transistor} for the analogy with a classical transistor and Fig.~\ref{fig:aqt} for a diagram of how an adiabatic quantum transistor would work under this analogy. Like a classical transistor this device is made to quantum compute via the application of an applied field to the device. Prior to the application of the field, the quantum information is localized to one side of the device, and the system is in its ground state. After the application of the field, the quantum information is on the opposite side of the device, but with a quantum circuit applied to this information, and the system remains in its ground state. The exact circuit applied depends on the microscopic details of the system---we propose that each such quantum transistor be made to implement a small fault-tolerant quantum circuit for a single encoded logical gate and be used to classically steer the quantum information across the device. During the application of the field, the system remains in its ground state, however the energy gap to the first excited state gets smaller. Thus the device functions as an instantiation of a universal adiabatic quantum computer~\cite{Aharonov:04a, Kempe:06a, Mizel:06a} with, however, the important differences that the ground state of the system can be degenerate (and thus the model could also be considered an open loop holonomic computation~\cite{Kult2006}), and, more importantly, that the information propagates spatially (thus the transistors are \textit{modular}.) This allows us to design clocked architectures that very closely resemble today's classical synchronous digital computers (see Fig.~\ref{fig:clock}). Here we introduce the details of adiabatic quantum transistors and present theoretical arguments and evidence from simulations that the energy gap in the system shrinks sufficiently slowly so as to allow adiabatic quantum computing. We further present arguments for how our construction can be made fault-tolerant. This results in an implementation of quantum computers in which precision gates, preparations, and measurements are replaced by sufficiently slow and smooth application of fields, and the degree to which one can suitably engineer a many-body quantum system.

Our starting point is the recent work that shows how it is possible to use \textit{piecewise} adiabatic deformation of a many-body interacting quantum system~\cite{Bacon:09a, Bacon:09b, Oreshkov:09a, Oreshkov:09b} to perform a quantum computation. In Ref.~\cite{Bacon:09b} we considered a certain many-body interacting quantum system in its ground state whose Hamiltonian we adiabatically deformed by turning on a strong field at the border of the device and then slowly propagating this field across the material;  this propagates quantum information through the device in a piecewise fashion, riding the front of the applied field. A major drawback of this method is that it requires one to precisely control a microscopic applied field that is turned on in a step-by-step process. This motivates the following question: if we take the construction in~\cite{Bacon:09b} and, rather than turning on the field in a piecewise fashion, we instead turn on the field simultaneously across the entire device, then does the device still perform the desired quantum computation?  Here we answer this question affirmatively and show how it leads to our adiabatic quantum transistor.

The many-body quantum system with the remarkable properties given in~\cite{Bacon:09b} is described by a twisted version of the Hamiltonian associated with cluster states. Cluster states are the entangled states originally used to perform measurement-based quantum computation~\cite{Raussendorf:01a, Briegel:01a}. To define the Hamiltonian of our system, we need a graph (with vertices $V$ and edge set $E$) where each vertex, $v$, is labeled by an angle $\theta_v$ and is associated with a qubit. We define the twisted cluster-state Hamiltonian to be
\begin{align}
\label{eq:hamc}
	H_C = - \sum_{v \in V} \bigl(\cos(\theta_v) [X]_v + \sin(\theta_v) [Y]_v\bigr)\!
	\prod_{(v,w) \in E} [Z]_w \,, 
\end{align}
where $[P]_v$ denotes the operator $P$ acting on the qubit at vertex $v$, $X$, $Y$, and $Z$ are Pauli operators, and we choose units so that the coupling constant is unity. The ground state of this Hamiltonian is a cluster state in a locally rotated frame~\cite{Bacon:09b}. Note that the above Hamiltonian contains interactions that are physically unrealistic because they involve more than two qubits. Indeed, the complexity of the interaction increases with the degree of the graph. Fortunately, we need only consider graphs with degree at most 3 (corresponding to four-body interactions on a honeycomb lattice graph), and it turns out that we can use perturbation theory gadgets to obtain an effective Hamiltonian of this form inside the low-energy sector of a Hamiltonian containing only two-qubit interactions~\cite{Bartlett:06a, Oliveira:08a, Bacon:09a} (see Sec.~\ref{sec:gadgets}). We will assume here the Hamiltonian in Eq.~(\ref{eq:hamc}) and return to implementation in two-qubit interactions later. 

\begin{figure}[t]
\begin{center}
\includegraphics[width=3.5in]{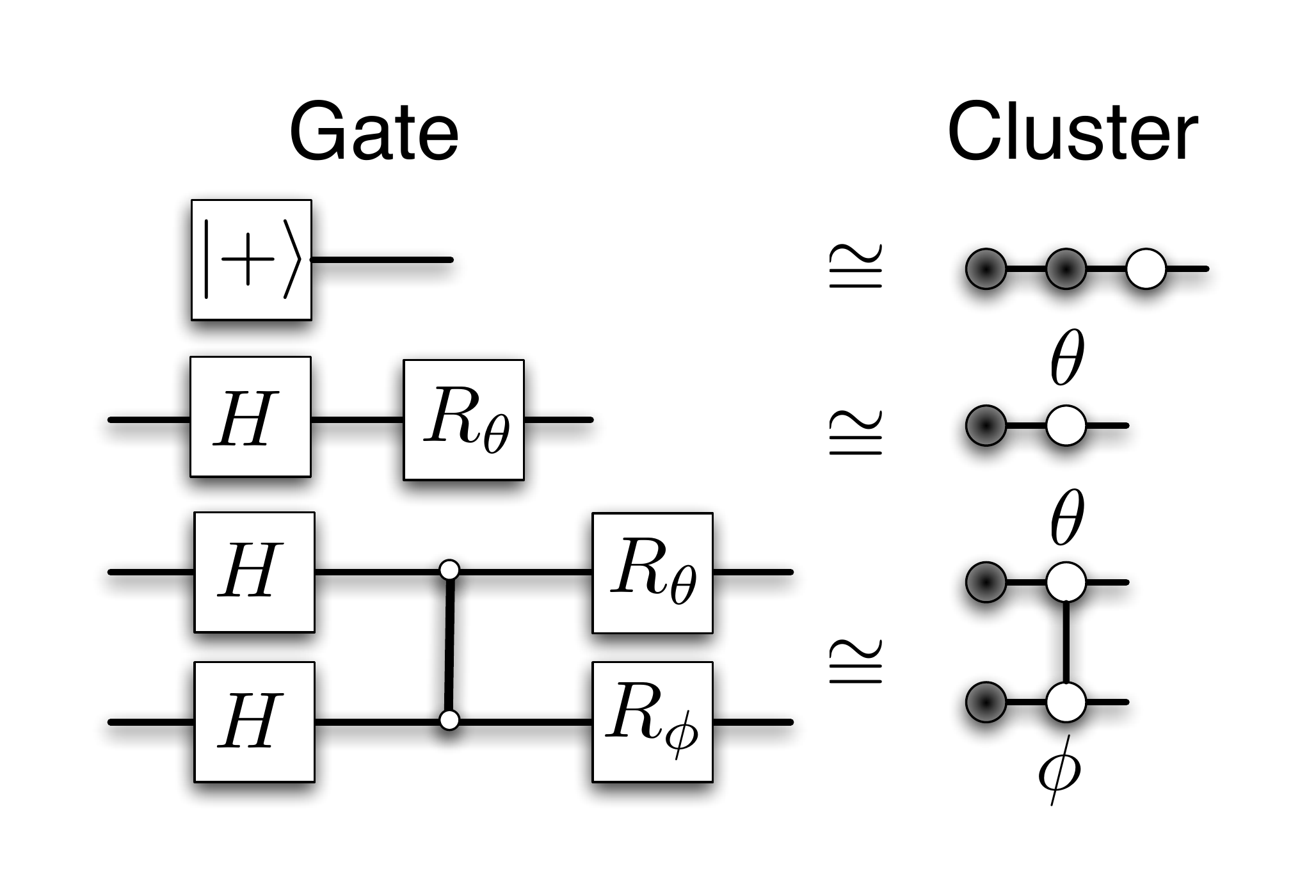}
\caption{Dictionary for translating a quantum circuit into the cluster state Hamiltonian (see one-way QC~\cite{Raussendorf:01a}.)  First, express the quantum circuit in the universal gate set above where $R_\theta=\exp(-i \theta Z/2) $, $H$ is a Hadamard, and the two-qubit gate is a controlled-phase gate, along with preparations of the $+1$ eigenvalue of $X$, $|+\rangle$. Then convert this circuit into a graph and corresponding twisted cluster-state Hamiltonian by converting the circuit graph to twisted cluster state graph using the diagramed dictionary.  For each of the unitary operators above, the corresponding gadget is labeled by its input (shaded) and output (unshaded) qubits. To construct the graph from the gate circuit, replace each unitary gate with the corresponding graph, merging output (unshaded) and input (shaded) qubits.  For the non-unitary preparation of $|+\rangle$ the inputs are never merged, but the output is merged with any input that this circuit element proceeds.  The end result of this construction will be a labeled (by the angles in the $R_\theta$ gates) graph corresponding to a twisted cluster-state Hamiltonian.  Output qubits will not have been merged and will be in the location of the quantum information after the adiabatic procedure described in the text has been completed.} \label{fig:recipe}
\end{center}
\end{figure}

The results in Ref.~\cite{Bacon:09b} show that one can take a quantum circuit and, using the recipe shown in Fig.~\ref{fig:recipe}, map it to a twisted cluster-state Hamiltonian with the property that adiabatically turning on fields along the $x$ direction ($-X_v$ on vertex $v$) and turning off relevant non-commuting terms in $H_C$ pushes information through the device in such a way as to implement the computation defined by the quantum circuit. In the analysis of Ref.~\cite{Bacon:09b} one can separate out the rigorous argument that the computation is performed from whether this procedure can be performed adiabatically. Thus, if we consider the case of turning on all of the fields simultaneously rather than piecewise, we see that our analysis of the fact that a quantum computation is performed carries directly over from Ref.~\cite{Bacon:09b}. By contrast, although the energy gap in the computation is independent of the size of the computation in the piecewise construction of~\cite{Bacon:09b}, when one turns on the field at all locations simultaneously this will no longer be true. Therefore, in order to show that turning on the field over the entire system still causes our device to execute the desired computation, we must show that the inverse of the energy gap between the ground state and the first excited states grows at most polynomially in the size of the quantum circuit being enacted. If indeed the inverse gap grows this slowly then the adiabatic theorem (for example~\cite{Schaller:06a}) guarantees that the system will remain in its ground state throughout the evolution conditional on turning on the field over a time scale polynomial in the size of the quantum circuit.

To summarize, quantum circuits correspond, via Fig.~\ref{fig:recipe}, to twisted cluster-state Hamiltonians, and the minimal spectral gap of such a Hamiltonian determines how long it takes, via adiabatic deformation, to achieve a high-fidelity implementation of the circuit. 

\subsection{A One Dimensional Wire} \label{sec:1d}

We begin by analyzing the simplest instantiation of our model: $n$ qubits on a line with no twists in the Hamiltonian, $\theta_v=0$ for all $v \in V$. In this case we evolve the system according to the time varying evolution
\begin{align}\label{eq:1d}
	H(t)= &- f(s) \biggl(-[Z]_{n-1} [X]_{n}+\sum_{i=1}^{n-2} [Z]_{i} [X]_{i+1} [Z]_{i+2} 
		\biggr) \nonumber \\
	 &\quad - g(s) \sum_{i=1}^{n-1} [X]_i
\end{align}
where $f$ and $g$ are suitably smoothly varying envelopes which satisfy $f(0)=g(1)=1$ and $f(1)=g(0)=1$, and $s=\frac{t}{T}$ is a scaled time. For simplicity, in this subsection, we will assume $f(s)=1-s=1-g(s)$. Note that we have also turned off the cluster state term in Eq.~(\ref{eq:hamc}) corresponding to the first qubit. This implies that the ground state of the Hamiltonian is two-fold degenerate and thus encodes a qubit~\cite{Bacon:09b}. Initially this logical qubit is localized on the first two physical qubits of the chain. At the end of the adiabatic evolution the qubit will be at the end of the chain on the last qubit, with a Hadamard gate applied to the qubit if the chain has even length. We can now ask the question, what is the minimum energy gap for the above evolution, and how does it scale with the length of the chain? If the energy gap scales as an inverse polynomial in the length of the chain, then the adiabatic theorem tells us that propagating the qubit down this length $n$ chain can be done efficiently (i.e., in a time polynomial in $n$).

To answer this question, we use the equivalence of this model to two uncoupled transverse Ising models~\cite{Doherty:09a} (see also~\cite{Pachos:04a}). These systems can be exactly diagonalized by a transformation of the spin model to a model with non-interacting fermions. Following the work of Doherty and Bartlett~\cite{Doherty:09a}, but with additional attention to the boundary terms, suppose that one defines a code with the following $(n-1)$ encoded Pauli operators
\begin{align}\label{eq:x}
	\bar{X}_j = 
	\left\{\begin{array}{ll}
		[X]_{1}[X]_{3} \cdots [X]_{j} & \mbox{if $j$  is odd}, \\
		{[}X]_{2}[X]_{4} \cdots [X]_{j} & \mbox{if $j$  is even} \,.
	\end{array}\right.
\end{align} 
and
\begin{align}\label{eq:z}
	\bar{Z}_j = 
	\left\{\begin{array}{ll}
		[Z]_j [X]_{j+1} [Z]_{j+2} & \mbox{if $j < n-1$}, \\
		{[}Z]_{n-1} [X]_{n} & \mbox{if  $j = n-1$} \,. 
	\end{array}\right.
\end{align}
Then, given these encoded operators, we can then express the Hamiltonian in Eq.~(\ref{eq:1d}) as the union of two uncoupled transverse Ising models acting separately on the even and odd qubits, with an additional boundary term:
\begin{align}
	H(s)=-(1-s) \sum_{i=1}^{n-1} \bar{Z}_i - s \bigg(\bar{X}_1+\bar{X}_2 + \sum_{i=0}^{n-3} \bar{X}_i \bar{X}_{i+2}\bigg) \,.
\end{align}

Using standard techniques for diagonalizing such Hamiltonians~\cite{Lieb:61a} (and the trick of adding an extra qubit to make the system have quadratic fermion operators, see Appendix~\ref{sec:lieb}), this Hamiltonian can be shown to be equivalent to a system of non-interacting fermions with
\begin{align}
	H_l(s) & = \sum_{k=1}^l \omega_k(s) \left (\eta_k^\dagger \eta^{\vphantom{\dagger}}_k-\frac{1}{2} \right) 
		\nonumber \\
	 \omega_k(s)^2 & = 4\left(1-2s(1-s)\Bigl[1-\cos\bigl(\tfrac{k \pi}{l+1}\bigr)\Bigr] \right)
\end{align}
where $\eta_k^\dagger$ is the creation operator for the $k$th fermion (see Appendix~\ref{sec:lieb}) and $l=\bigl\lfloor \frac{n}{2} \bigr\rfloor$ or $l=\bigl\lfloor \frac{n+1}{2} \bigr\rfloor$ depending on whether one is considering the even or odd chain. Note that there is no $k=0$ energy level. From this equation one sees that the minimum energy gap occurs at $s=\frac{1}{2}$ where $\omega_k(1/2)=2 \cos \bigl( \frac{k \pi}{2(l+1)} \bigr)$ and is of order $O\bigl(\frac{1}{n}\bigr)$. Thus, since the minimum gap scales inversely as a polynomial in the length of this one-dimensional system, we see that if we turn the field on in a time polynomial in this length, then with high probability the quantum information will propagate from one end of the system to the other end---with a possible Hadamard gate applied, depending on whether $n$ is odd or even.

\subsection{Single Qubit Quantum Circuits} \label{sec:1dgates}

Having demonstrated rigorously that the scaling for a quantum wire has a gap which scales inversely with the length of the wire, we now examine what happens when we apply other single qubit gates by using twisted Hamiltonians with varying $\theta$'s. Here we give strong numerical evidence that the gap scales inversely as a polynomial through the use of a matrix product state algorithm. To simplify our study of the one dimensional twisted Hamiltonian, it is convenient to note that this model has a duality. In particular, for a twisted Hamiltonian with angles $\theta_i$ where $1 \leq i \leq n-2$ and $\theta_i$ is the angle associated with the $(i+1)$st qubit, it can be shown that if the angles obey the symmetry condition $\theta_i=\theta_{n-i-1}$ for all $i$ and additionally there is only one minimum in this model (which is true numerically for small systems), then the minimum energy gap occurs at the midpoint, $s=1/2$. 

To see this, we proceed as follows. First we define a shorthand for the operators which are rotated combinations of $X$ and $Y$ Pauli operators, namely,
\begin{align}
	[\theta_k]_i = \cos(\theta_k) [X]_i + \sin(\theta_k) [Y]_i \,.
\end{align}
Now consider the twisted one-dimensional cluster Hamiltonian with a transverse field,
\begin{align}
	H(s) = & -(1-s) \Biggl( \sum_{i=1}^{n-2} [Z]_{i}[\theta_i]_{i+1} [Z]_{i+2} 
		+ [Z]_{n-1} [X]_{n}\Biggr) \nonumber \\  
	& \qquad- s \sum_{i=1}^{n-1} [X]_i.
\end{align}
Define the following unitary operator:
\begin{align}
	U = S \Biggl( [Z]_1 \prod_{i=2}^n [X]_i \Biggr) 
	\Biggl( \prod_{i=1}^{n-1} [C_Z]_{i,i+1} \Biggr) 
	\Biggl( \prod_{i=1}^{n-2} [R(-\theta_i)]_{i+1} \Biggr)
\end{align}
where $R(\theta)=\exp(-i Z \theta/2)$, $[C_Z]_{i,j}$ is the controlled-phase gate acting between qubits $i$ and $j$, and $S$ is the gate which inverts the chain about its middle, swapping the $1$st and $n$th, $2$nd and $(n-1)$st, etc.\ qubits. Then one can check that
\begin{align}
	UH(s)U^\dagger = & - s \biggl( \sum_{i=1}^{n-2} [Z]_{i}[\theta_{n-i-1}]_{i+1} [Z]_{i+2} \nonumber \\
	& \quad - [Z]_{n-1} [X]_{n}\biggr) - (1-s) \sum_{i=1}^{n-1} [X]_i.
\end{align}
Thus, in the case that $\theta_i=\theta_{n-i-1}$ for $1 \leq i \leq n-2$, $H(s)$ has the same spectrum as $H(1-s)$ and in particular if there is only one quantum phase transition (meaning, only one value of $s$ for which the spectral gap above the degenerate ground space collapses to zero in the thermodynamic limit), then it occurs when $s=1-s$ or $s=\frac{1}{2}$. 

\begin{figure}
\begin{center}
\includegraphics[width=3.5in]{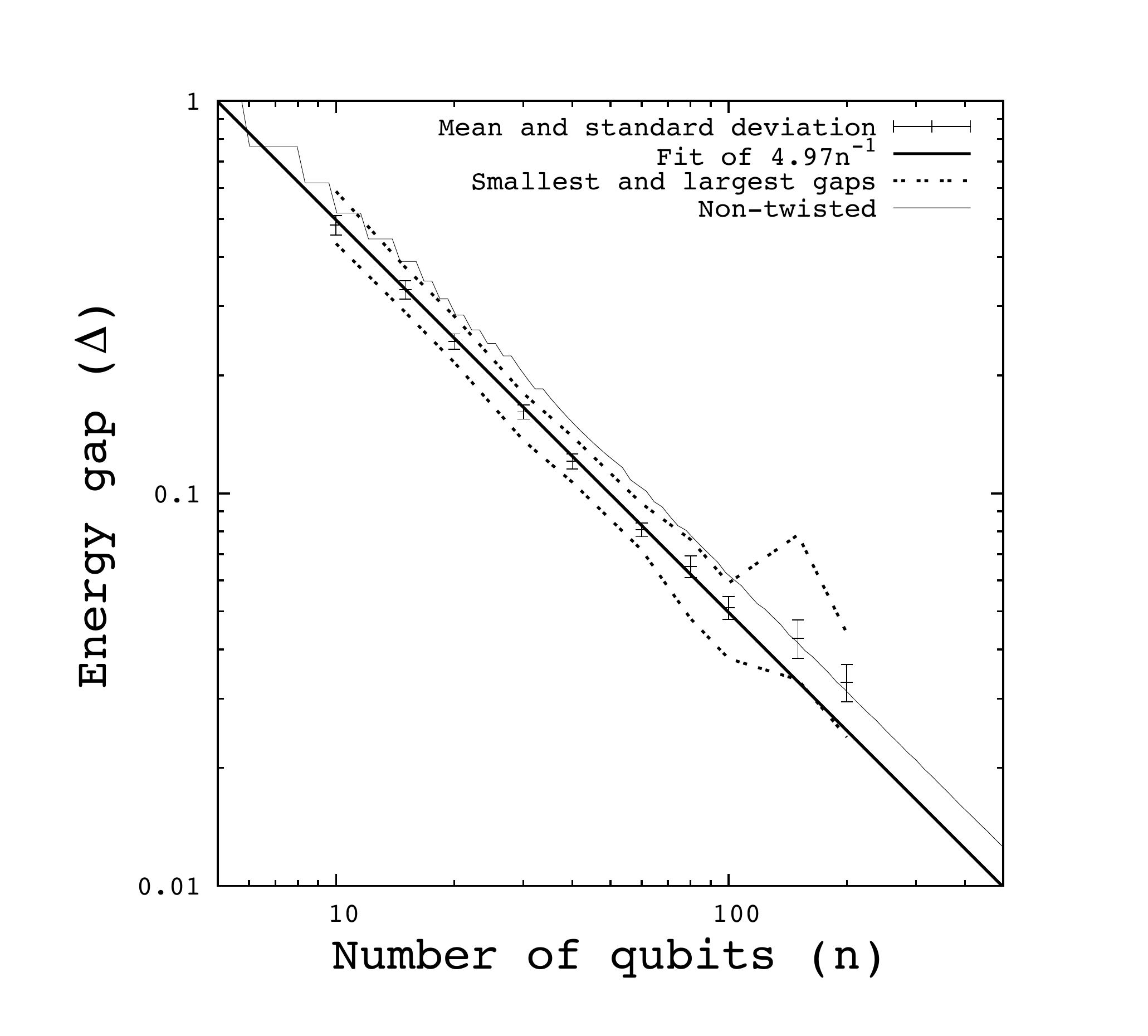}
\caption{Energy gap versus number of qubits for a one-dimensional twisted Hamiltonian with random angles. The duality condition has been enforced so that the minimum is at $s=\frac{1}{2}$ (see text.)  We took $200$ samples with uniformly distributed random angles. We also plot the smallest and largest energy gaps that were obtained for a given number of qubits (dotted lines) as well as the case with no rotated angles (which we showed could be computed exactly) (non-bold line.)   The data are well fit by $(4.97 \pm 0.02) n^{-1}$.} \label{fig:1dgap}
\end{center} 
\end{figure}

Given this, we used the technique of \emph{matrix product states}~\cite{Verstraete:04a, White:92a, Schollock:05a} to calculate the energy gap for systems having up to $200$ qubits with angles $\theta_i$ chosen uniformly from $[0,2 \pi)$ satisfying the aforementioned symmetry condition;  for each system size that was examined, we performed 200 simulation runs, each with a different choice of angles. Fig. \ref{fig:1dgap} presents the resulting data, from which we find that the energy gap data are well fit by a function which scales as $O(n^{-1})$, thus indicating that the twisting of the Hamiltonian does not quantitatively change the scaling of the energy gap with the length of the single qubit circuit being implemented. 

\subsection{Multi-qubit Quantum Circuits} \label{sec:2d}

Finally we turn to the question of the energy gap in the case where the circuits involve two or more qubits. Here we note that if we take a square lattice with appropriate boundary conditions, then with the untwisted Hamiltonian model is equivalent to a quantum compass model restricted to a certain symmetry sector. Under this symmetry restriction, the minimal gap occurs at $s=\frac{1}{2}$, where the compass model has equally competing interactions. Numerical evidence from exact diagonalization on a square lattice for this model presented in Ref.~\cite{Dorier:05a} demonstrates that the minimal energy gap for this model with the symmetry restriction scales as $~1/n$ where $n$ is the size of (i.e., number of qubits in) the square lattice. Thus the real question is what happens to the energy gap for more general circuits.

We consider a model on a square lattice with the full twisted cluster state Hamiltonian turned on interpolating to a Hamiltonian with applied fields on across the entire device. Label the qubits on the square grid by $(i,j)$. The Hamiltonian we consider is
\begin{align}\label{eq:h2d}
	H =&  (1-s) H_C - s \sum_{i,j=1}^L [X]_{(i,j)} 
\end{align}
where $L$ is the length of the lattice and we define $[P]_{(i,j)}$s with values of $i$ or $j$ lying outside the lattice as identity operators to deal with boundary terms. Here, $H_C$ is the twisted cluster Hamiltonian of Eq.~(\ref{eq:hamc}) for a square lattice with some particular choice of angles $\theta_{(i,j)}$ for each vertex $(i,j)$ in the lattice. We suppress the dependence on the $\theta_{(i,j)}$ to avoid notational clutter.

If we apply a unitary operator $U_x$ consisting of a controlled-phase gate between all $x$-coordinate neighbors, $[C_Z]_{(i,j),(i+1,j)}$, to the Hamiltonian in Eq.~(\ref{eq:h2d}), then the cluster state Hamiltonian is turned into a $y$-direction striped twisted Hamiltonian and the applied field turns into a $x$-direction striped untwisted Hamiltonian. Explicitly, this transformed Hamiltonian (which will have the same spectrum) is given by
\begin{align}\label{eq:hy}
	U^{\vphantom{\dagger}}_x H U_x^\dagger = & -\sum_{i,j=1}^L  (1-s)  [\theta_{(i,j)}]_{(i,j)} [Z]_{(i+1,j)} [Z]_{(i-1,j)} \nonumber \\
	& \quad + s [Z]_{(i,j-1)} [X]_{(i,j)} [Z]_{(i,j+1)} \,.
\end{align} 
If one instead applies $U_y$, consisting of controlled-phase gates between all $y$-coordinate neighbors, $[C_Z]_{(i,j),(i,j+1)}$ to the Hamiltonian in Eq.~(\ref{eq:h2d}), then we see that this simply swaps the direction of the stripes:
\begin{align}\label{eq:hx}
	U^{\vphantom{\dagger}}_y H U_y^\dagger = & -\sum_{i,j=1}^L  (1-s) [\theta_{(i,j)}]_{(i,j)}[Z]_{(i,j-1)} [Z]_{(i,j+1)}  \nonumber \\
	& \quad +s [Z]_{(i-1,j)} [X]_{(i,j)} [Z]_{(i+1,j)} \,.
\end{align} 
If one swaps the qubits about the line $x=y$ in Eq.~(\ref{eq:hx}), we obtain the striped Hamiltonian in Eq.~(\ref{eq:hy}), but with $\theta_{(i,j)}$ replaced by $\theta_{(j,i)}$ and $s$ replaced by $1-s$. Thus if we enforce $\theta_{(i,j)}=\theta_{(j,i)}$ we obtain a duality, and as in the previous subsection there will be a minimum at $s=\frac{1}{2}$. 

\begin{figure}
\begin{center}
\includegraphics[width=3.5in]{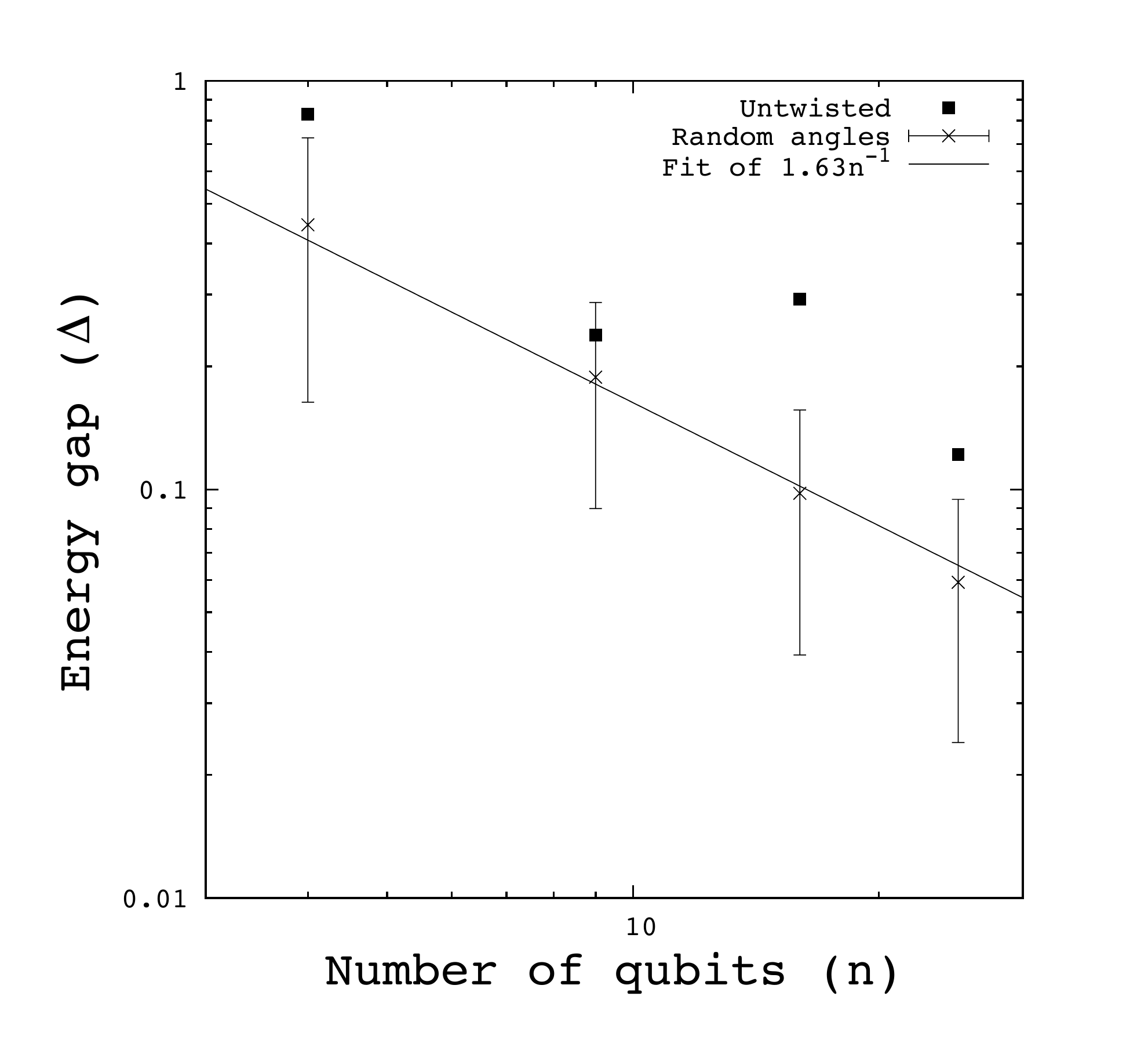}
\caption{Energy gap versus number of qubits for the two-dimensional twisted Hamiltonian with random angles and the duality condition enforced (see text.)  The initial Hamiltonian was a rotated cluster Hamiltonian for a square lattice (Eq.~(\ref{eq:hamc})) and the final Hamiltonian was an $X$ field over the entire device. All data were obtained using exact diagonalization with accuracy $10^{-8}$, with the exception of the random angle data for $n=25$ which were obtained using a one dimensional matrix product algorithm wrapped onto the two-dimensional lattice and have accuracy $10^{-3}$. The data are well fit by $(1.63 \pm 0.12) n^{-1}$. We also show the untwisted energy gap for comparison. The untwisted energy gap does not appear to be a power law because there is an even-odd lattice effect: examination of non-square lattices gives us more confidence that the untwisted case is indeed a power law.} \label{fig:2dgap}
\end{center}
\end{figure}

For systems that obey the duality condition $\theta_{(i,j)} = \theta_{(j,i)}$ we have investigated the size of the energy gap using exact diagonalization as well as a matrix product state approach. The results of these simulations are plotted in Fig~\ref{fig:2dgap}. While we can only obtain weaker evidence of an inverse polynomial for this two-dimensional system, the data are at least consistent with this hypothesis.

\section{Fault-tolerance of the Adiabatic Quantum Transistor} \label{sec:ft}

Having given strong evidence that our model can efficiently produce a desired quantum circuit when all evolutions are error free, we now turn to the question of how this model will behave in realistic settings where the system is coupled to an environment. Here we argue that if one uses standard circuit constructions for fault-tolerant QC~\cite{Aharonov:97a, Knill:98a, Knill:98b}, this will be good enough to ensure robust quantum computation in the model. In particular we can consider a model in which each modular adiabatic quantum transistor is configured to perform an encoded quantum gate, state preparation, or measurement from a fault-tolerant quantum circuit construction, including the error correcting step. We will argue that if one does this, then the standard analysis of the success probability of such constructions carries over to our model. Note that fault-tolerant QC requires expunging entropy (usually via measurement), but this can always be placed at the end of the fault-tolerant block~\cite{Paz-Silva:10a}. Thus we envision here a model in which the final measurements in error correcting circuits are implemented after each evolution of a quantum transistor. This can be done by either directly measuring the relevant qubits, or via an adiabatic amplifier that dissipates energy by the natural relaxation of the system, as described in Sec.~\ref{sec:amp}.

First note the following positive result about using each quantum transistor as a fault-tolerant circuit gadget. If one is executing a quantum algorithm with $L$ gates, then each of these gates needs to be executed with accuracy $\epsilon=\epsilon_0 L^{-1}$. To do this using fault-tolerant gates below the fault-tolerant threshold requires~\cite{Aharonov:97a, Knill:98a, Knill:98b}  that the gadgets have circuits of size $O\bigl(\log^c \epsilon^{-1}\bigr)$ for a constant $c$. Thus if the energy gap in an adiabatic gate shrinks as an inverse polynomial, the energy gap in the individual fault-tolerant gadget will shrink as $O\bigl( (\log^k L)^{-1}\bigr)$ for a constant $k$. Because each individual transistor executes a single encoded gate, this means that each adiabatic evolution need only be polylogarithmically longer when executing larger and larger quantum algorithms.

Second, and equally important, is the fact that a spectral gap \emph{need only hold for some universal set of encoded gates}. In other words, we do not need to have a spectral gap for all possible Hamiltonians of the type in Eq.~\ref{eq:hy}, we only need a gap for some particular set of logical gates. For example, if one could prove that logical Hadamard, CPHASE and $\pi/8$ gates have an appropriately large gap, that would be sufficient for any algorithm in our scheme to have a sufficiently large gap, since it would be comprised of solely these gates. Having a gap for only a handful of specific models is a tremendously weaker requirement than having a gap for all twisted cluster Hamiltonians. We expect that, at the very least, substantial numerical evidence could be gathered in support of a gap for a universal set, and indeed our results in Fig.~\ref{fig:2dgap} already constitute some measure of support for this statement. Proving a gap for a universal set of logical gates remains an important open problem related to this work. 

\subsection{An Unrealistic But Illuminating Error Model}

We now turn to the question of whether fault-tolerant gadgets will act to overcome errors in the adiabatic transistor. We begin by considering the following simplified error model: assume that at the beginning of the computation the system is not in the ground state of $H_C$, but rather it is in one of the excited states of $H_C$. Note that since the terms of $H_C$ commute (as expressed in Eq.~\ref{eq:hamc}), excited states of $H_C$ can be labeled by the list of eigenvalues of each of the terms, where the eigenvalue of each term is either $+1$ or $-1$. If one carries out the analysis described in Ref.~\cite{Bacon:09b} regarding the computation that is performed if one starts in such an excited state, then one sees that for each of the terms in $H_C$ whose initial eigenvalue is $+1$ there is a corresponding Pauli error in the circuit---in other words, errors that take the form of starting in an excited state of $H_C$ map directly onto errors in the circuit model. For example, consider the one-dimensional case with no rotated angles; if the initial state is in the $+1$ eigenspace of some term $- [Z]_{i-1}[X]_i [Z]_{i+1}$ in the middle of the chain, then this corresponds to a Pauli $Z$ error in the circuit model.

Furthermore, note that in this model errors that are local on the physical qubits of $H_C$ map to errors that are local in the circuit since all terms in $H_C$ are localized on a few qubits---so for example, applying $[X]_i$ to the system flips only two eigenvalues of $H_C$ in the untwisted case.  By the linearity of quantum mechanics, we can extend this argument to initial preparations that are mixtures and coherent superpositions of such errors. Thus if we take our initial system and expose it to an environment for an amount of time proportional to the size of the quantum circuit we are about to enact, then subsequent perfect adiabatic evolution will produce errors in the quantum circuit model that are commensurate with a local independent error model. (Note that this model does have temporally correlated errors over a few qubits, but this does not affect the existence of fault-tolerant methods for overcoming these errors~\cite{Aharonov:97a}.)

\subsection{Errors On A Quantum Wire}

Of course, this model in which errors appear only at the beginning of the computation is a fiction. More generally we may consider a model in which errors occur throughout the adiabatic evolution. Consider such an evolution for the single quantum wire with an untwisted Hamiltonian. As we detailed above, this model can be mapped into two transverse-field Ising models. Any linear operator, and hence any error, on this model can be expressed as one of three types of errors: errors that change the energy of the system, errors that act on the degeneracy (corresponding to the encoded qubit), and a combination of these two errors. We will show that each of the first two types of errors can be mapped to an independent error model on the quantum circuit corresponding to the quantum wire in an independent manner, thus taking care of the third type of error. Finally note that we use the term \textit{error} in a quantum error correcting sense to denote an error operator, even when this arises from coherent error sources. For example our results deal with the errors in the Hamiltonian description resulting from a perturbing interaction on the system $H^\prime=H+\lambda V$ for small $\lambda$ as well as for couplings of our system to an environment, though the argument for these errors is slightly different (see Sec.~\ref{sec:static} for a discussion of this point.)

\subsubsection{Errors that change the energy of the system}

We first consider errors that change the energy of the system. The energy spectrum calculated in Sec.~\ref{sec:1d} has low-lying excited states and thus we can reasonably expect excitations into these levels. Assuming detailed balance in our error rates (which tends to be the case for weak coupling errors that change the energy of a system~\cite{Childs:01a}) we consider a model in which the rate of errors that change energy by $\Delta E$ scales as $p_0 \mathrm{e}^{- \frac{\Delta E}{k_B T}}$ where $p_0$ is a bare error rate, $k_B$ is the Boltzmann constant, and $T$ is the temperature of the environment. 

Consider working at a temperature much less than the bare energy gap of the system at $s=0$, which is $\Delta =2$. In this case the only fermionic excitations that will occur with high probability are those for which $\omega(s) < k_BT$. This condition implies that only a constant fraction of fermions will be created in such an error model throughout the adiabatic evolution. Furthermore, note that because the adiabatic evolution preserves the energy levels, it will therefore proceed to drag the system along the excited energy level. We see, therefore, that errors that excite a constant fraction of fermions can be mapped back to excitations at the beginning of the adiabatic evolution; as we have already shown, this is not a problem for standard quantum-error-correction procedures. 

\begin{figure}
\begin{center}
\includegraphics[width=3.5in]{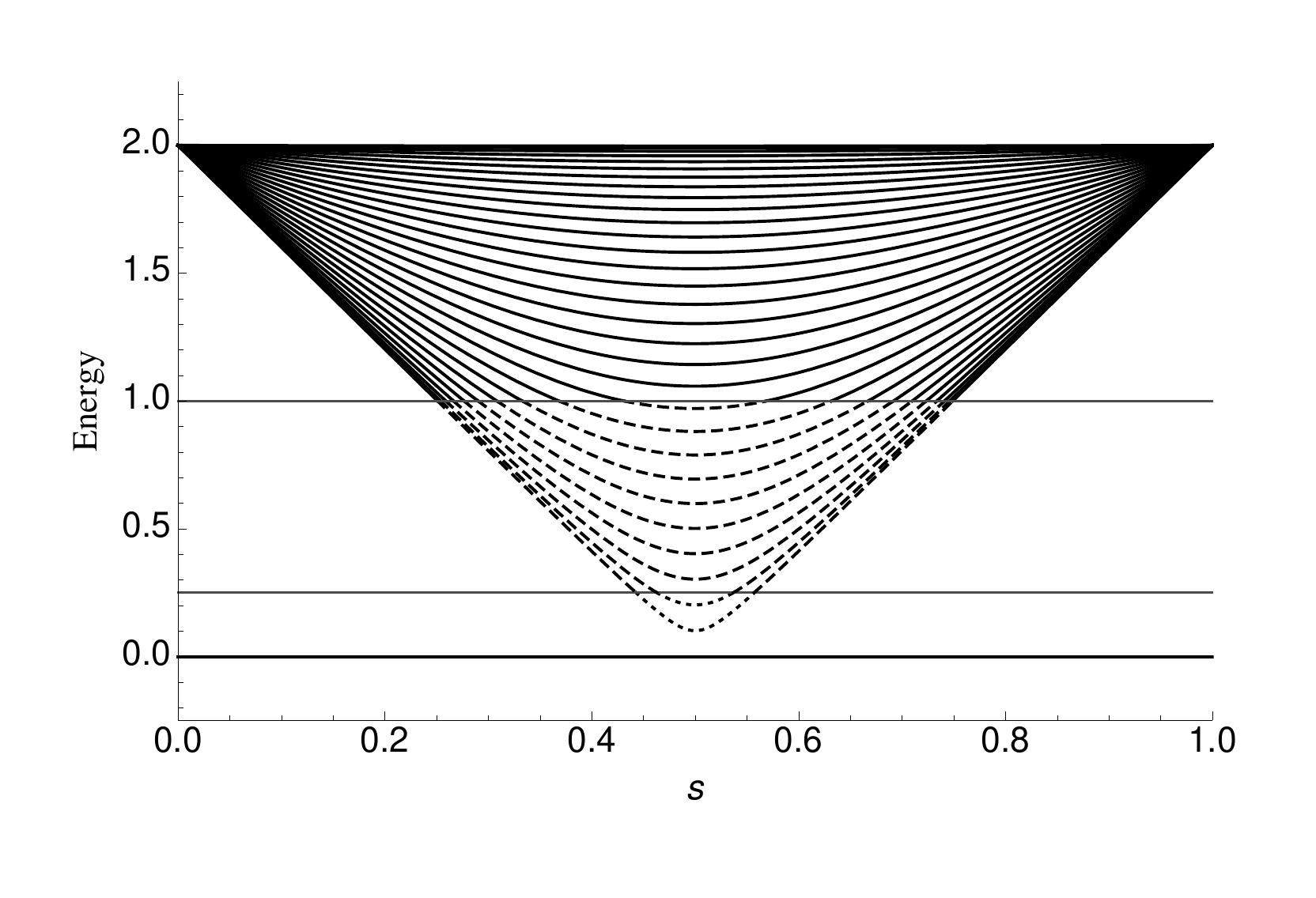}
\caption{When one exactly solves the one-dimensional wire, the result is that the wire behaves like a collection of free fermions. Here we show the energies for creating each of these fermions in a chain of length $80$ (thus all the energy levels are sums of these energies) as a function of the scaled time $s$. See Eq.~(5) in Appendix~\ref{sec:lieb}. For high energy excitations, if the environment is cool enough, then these energy levels are essentially never excited (solid lines.) At energies comparable to the temperature of the environment, errors will no longer be suppressed by a Boltzmann factor (dashed lines.) At the lowest energies, the interaction strength between the environment and the system is small enough that the error model will not be governed by detailed balance. However, assuming that these energy levels are changed by any process whatsoever is no worse than our assumption that \emph{all} errors below the temperature cutoff lead to real errors.} \label{fig:figure7}
\end{center}
\end{figure}

Finally we note a technical condition: we have assumed a detailed balance error model, but for the lowest lying energy levels, where the strength of the interaction between the system and environment is greater than the energy gap, this condition might not hold. However, these energy levels are, under our temperature assumption, already assumed to be in error, and therefore covered by our argument (i.e., they are at most a constant fraction of the error, the fraction being directly related to the ratio of the energy gap and the perturbation energy strength.) See Fig.~\ref{fig:figure7} for a graphic depicting the relevant energy levels and the partition of these errors into different categories in our argument. 

To summarize: under the reasonable assumption that high energy excitations are suppressed by a Boltzmann factor, we see that errors in the quantum wire that change the energy of the system result in independent errors on the quantum circuit being enacted. 

\subsubsection{Errors on encoded quantum information}

The situation for errors that do not change energy but instead act on the degeneracy of the quantum wire is slightly more complex. To see what happens, we consider the two logical operators encoded in the degenerate ground state of the quantum wire, which for a chain of odd length $n$ are given by
\begin{equation}
\bar{X}_n= \prod_{i=1}^{\frac{n+1}{2}} [X]_{2i-1} \quad {\rm and} \quad \bar{Z}_n= [Z]_{n} \prod_{i=1}^{\frac{n-1}{2}} [X]_{2i}.
\end{equation}
The first $n-1$ operators are the encoded operators of the code described in Sec.~\ref{sec:1d}.
Note that these two logical operators commute with the entire Hamiltonian during the adiabatic evolution. At the beginning (end) of the computation this information is localized onto the start (end) of the quantum wire. Thus during both of these times the quantum information in the wire is susceptible to decoherence from local errors. For example undesired or imprecisely controlled terms in our initial Hamiltonian will be able to act non-trivially on the initial quantum information. However, during the middle of the computation, we will argue that the information in the degeneracy is protected, and furthermore that we can adjust the adiabatic schedule in such a way that the system only spends a constant amount of time during which the quantum information is exposed to local decoherence at the beginning and end of the adiabatic evolution.

In particular consider the logical $\bar{X}_n$. By expanding this logical operator in terms of the underlying physical operators, we see that this error requires a combination of $O(n)$ local $[X]_i$ errors on odd qubits. When we map these errors over to the transverse Ising model, these are errors on the Ising model, but are now two-qubit terms like $\bar{X}_i \bar{X}_{i+1}$ for $i \geq 2$ or $\bar{X}_1$. Individually these errors can change the energy of the system (and we have previously argued that such errors can be dealt with), but they can also have an effect which keeps the energy constant. To evaluate this effect, for example, for information at the end of the computation, we can evaluate the portion of the error amplitude which preserves the vacuum of the transverse Ising model: $e =\prod_{i=0}^{\frac{n}{2}} \langle v| \bar{X}_i \bar{X}_{i+1} |v\rangle$ where $|v\rangle$ is the vacuum (the single $\bar{X}_1$ error can be thought of as a two-qubit $X$ error on the fictitious extra qubit.)

For the transverse Ising model with periodic boundary conditions in the thermodynamic limit, the nearest neighbor $XX$ correlations are given by~\cite{Pfeuty:70a} 
\begin{align}
	\langle v| [X]_i [X]_{i+1} |v\rangle = 
		\frac{1}{\pi} \int_{0}^{\pi} \Omega^{-1}(k)\bigl[\cos(k)+\alpha \bigr]\,{\rm d}k \,,
\end{align}
where
\begin{align}
	\Omega(k)=\sqrt{1+\alpha^2+2 \alpha \cos k}
\end{align}
and, to relate back to the quantum wire model, $\alpha=\frac{s}{1-s}$. This can be transformed into
\begin{align}
	\langle v| [X]_i [X]_{i+1} |v\rangle = & \frac{1}{\pi s} (2s - 1) K\bigl[4 s (1-s)\bigr] \nonumber \\
	& \quad + \frac{1}{\pi s} E\bigl[4 s(1-s)\bigr] \,,
\end{align}
where $K$ and $E$ are the complete elliptic integrals of the first and second kind respectively. We are interested in the amplitude for the product of $n$ of these correlation functions, $\langle v| [X]_i [X]_{i+1} |v\rangle^{n/2}$. Series expanding this about $s=1$, we find 
\begin{align}
	\langle v| [X]_i [X]_{i+1} |v\rangle^{n/2} = & \,
		1-\frac{1}{8} n (s-1)^2 + \frac{1}{4} n (s-1)^3 \nonumber \\
		&\quad + \biggl(\frac{2 n^2 - 53 n}{128}\biggr) (s-1)^4 \nonumber \\
		&\quad + O\bigl((s-1)^5\bigr) \,.
\end{align}
If $(1-s)<O\bigl(n^{-\frac{1}{2}-\epsilon}\bigr)$ for $\epsilon>0$, we see that this function vanishes as $n \rightarrow \infty$. 

Thus using correlation functions for the relevant degeneracy-preserving operation we see that the amplitude for the degeneracy-preserving error is exponentially small (as a function of $n$) except for $s$ near $s=1$ in a window of size $O(n^{-{\frac{1}{2}}-\epsilon})$ for $\epsilon>0$.  In and of itself this would imply that the model is in trouble: with a linear interpolation for this would imply spending $O(\sqrt{n})$ time in this region during which the quantum information can be decohered. However, note that the energy gap for this model is large during the beginning and end of the adiabatic path. Thus instead of using a linear interpolation as in Eq.~(\ref{eq:1d}) one can use an interpolation which spends only a constant amount of time in the window where these errors can affect the logical quantum information.

In particular if we maintain the adiabatic condition for each infinitesimal evolution of the system we can obtain a schedule for interpolating our Hamiltonian which spends significantly more time where the energy gap is smaller. In particular, consider the adiabatic condition for the one-dimensional quantum wire. As we show in the appendix, the energy spectrum for a wire of length $2l$ is given by Eq.~(\ref{eq:energy}). To maintain the adiabatic condition for the lowest energy level infinitesimally we must satisfy~\cite{Roland:02a}
\begin{align}
	\frac{{\rm d}s}{{\rm d} t} = \epsilon \omega_l^2(s) 
	\quad \Leftrightarrow \quad 
	{\rm d}t = \frac{{\rm d}s}{\epsilon \omega_l^2(s)} \,,
\end{align}
where $\epsilon$ is the accuracy that we desire for our adiabatic evolution relative to the ideal evolution. Performing this integration and enforcing $t=0$ at $s=0$ yields the solution
\begin{align}
	t =& \frac{1}{4 \epsilon} 
		\csc \Big( \frac{\pi}{l+1}\Big) \nonumber \\ 
		& \quad \bigg( \frac{\pi l}{2(l+1)}
		+ \tan^{-1} \Big[ (2 s -1) \cot\big( \tfrac{\pi}{2(l+1)} \big) \Big] 
		\bigg).
\end{align}
Inverting this equation for $s$ yields a schedule for adjusting $s$ as a function of time which runs for a total time of
\begin{align}
	T=  \frac{\pi l}{4 \epsilon(l+1)} \csc \Big(\frac{\pi}{l+1}\Big)
\end{align}
which scales linearly as $l$ for large $l$ (thus there is no Grover-type speedup using this schedule.)  The amount of time during which this schedule spends with $1-\eta \leq s \leq 1$ is 
\begin{align}
	t_\eta = \frac{T}{2 \epsilon} 
		\bigg( 1+ \frac{2(l+1)}{\pi l} 
		\tan^{-1} \Big[ (2\eta-1) \cot\big( \tfrac{\pi}{2(l+1)} \big) \Big]
		\bigg) \,.
\end{align}
For the case relevant to the degenerate error model where $\eta=l^{-1/2}$ this is bounded by $\frac{1}{\epsilon}$. Thus this schedule spends only a constant amount of time in the region where the logical output quantum information is susceptible to noise. 

In conclusion, for the single-qubit quantum wire case, we see that the quantum information in the degeneracy is vulnerable to degeneracy-preserving errors only at the beginning or end of the computation, and in a setting that is no worse than the standard circuit model with independent errors. We can adjust the adiabatic schedule so as to not spend much time in these regions. Actually, this is no different than how we would want to operate a quantum computer with or without adiabatic quantum transistors: to keep the effective error-corrected error rate down, one needs to spend a minimal amount of time letting the quantum information simply decohere. In our model this means that we need to move faster than linearly out of the region where the encoded quantum information is exposed to errors. We note that the above analysis works for both open system errors and also for (small) deviations in the Hamiltonian we implement, i.e., perturbative system errors.

\subsubsection{Static Errors} \label{sec:static}

In the previous subsection we argued that an adiabatic quantum transistor configured to act as a quantum wire would behave, given an appropriate adiabatic schedule, as a quantum circuit with an independent error model. In this argument we have explicitly used an argument that relied on detailed balance. Thus one might wonder how the argument works when there is no environment and the errors occur solely from a static perturbation on the wire. Here we argue that these errors also give rise to an independent error model, at least for a quantum wire with no gates. 

In particular consider a model in which a static term is added to our adiabatic evolution $H^\prime(s)= H(s)+\lambda V$ for the case of a quantum wire. We will assume that $V$ is a sum of local operators (for example, one- or two-qubit operators) and that the strength of the interaction is perturbative ($\lambda \ll 1$, since we have fixed to unity the energy scale used in the twisted cluster-state Hamiltonian and the final applied field). First note that the contribution of $V$ to splitting the degenerate ground state of the wire will be exponentially suppressed except at the beginning and end of the adiabatic evolution, as we have argued above. The reason for this is exactly the same: local operators do not act on this space except when multiplied together (except at the beginning and end of the adiabatic evolution). Thus a term like $\lambda V$ will act like a single-qubit error term only for the time we spend with the information localized on either end of the wire. Using the scheduling trick we have described previously, this means that it acts like an independent error (from a static term) at the beginning and end of the computation represented in the wire. Thus we see that the static perturbing term will not act on the degeneracy in a manner that is inconsistent with an independent error model. 

Moreover, many types of errors will hardly effect the degeneracy at all since it is known~\cite{Son2011} that the cluster state has symmetry-protected topological order~\cite{Schuch2011, Chen2013} and the degeneracy and spectral gap for such systems are known to be stable~\cite{Bravyi2010, Michalakis2011}. (See also Refs.~\cite{Else2012, Miyake2010} for other schemes for quantum computation in a symmetry-protected topologically ordered phase.)

Next we turn to the case in which there is a static perturbation, but where we now concern ourselves with the effect this has coming from portions of $V$ that do not act on the degeneracy. These perturbations have the effect of changing the energy levels, in terms of both the eigenvalues and the eigenvectors of the system. 

Consider first the effect of the perturbation on the eigenvectors of the system. We begin by recalling the argument that shows that the quantum wire Hamiltonian correctly transmits information in the absence of a static perturbation. Recall that the quantum wire Hamiltonian is given by
\begin{align}
	H(t)=&- f(s) \biggl(\Delta \sum_{i=1}^{n-2} [Z]_{i} [X]_{i+1} [Z]_{i+2} -\Delta [Z]_{n-1} [X]_{n}\biggr) \nonumber \\
	 &\quad - g(s)\Delta \sum_{i=1}^{n-1} [X]_i. \label{eq:wire2}
\end{align}
Initially we can define the logical operators $\bar{X}=[X]_1 [Z]_2$ and $\bar{Z}=[Z]_1$ and the system is in the $+1$ eigenstate of the vertex operators $[Z]_i [X]_{i+1} [Z]_{i+2}$ ($1 \leq i \leq n-2$) and $[Z]_{n-1}[X]_n$. By suitable multiplication of the encoded logical operators with these vertex operators, we can express the logical operators as a pattern of $[X]_i$ operators on qubits, for $i<n$, and a Pauli operator on the last qubit: $\bar{X} \equiv \left( \prod_{i=1, i~{\rm odd}}^{n-1} [X]_i   \right) [X]_n$, $\bar{Z} \equiv \left( \prod_{i=1,i~{\rm even}}^{n-1} [X]_i \right) [Z]_n$, where we have assumed for simplicity that $n$ is odd. At the end of the evolution we will be in the $+1$ eigenstate of all of the $[X]_i$, $1 \leq i \leq n-1$. This result then implies that the information will be correctly transmitted down the quantum wire since under this condition the logical operators are just bare Pauli operators on the last qubit. 

Suppose, on the other hand, that we are initially in a state in which we are in the $-1$ eigenvalue eigenspace of a single vertex operator. Then the above argument about the logical operators could still be applied, but the effect would be that one of the logical operators might acquire a phase. For example, it might end up that $\bar{X} \equiv -\left( \prod_{i=1, i~{\rm odd}}^{n-1} [X]_i   \right) [X]_n$, $\bar{Z} \equiv \left( \prod_{i=1,i~{\rm even}}^{n-1} [X]_i \right) [Z]_n$, which represents the quantum wire being faithful up to a $Z$ error. Generalizing, we see that initial states that are not in the $+1$ eigenstates translate into single-qubit errors on the wire (so that multiple vertex operators in $-1$ eigenstates represent multiple errors.) A similar argument applies to the end of the evolution: if we are are in $-1$ eigenstates of some final $[X]_i$ operator, this will correspond to a single-qubit error on the wire. Note that it is possible for the errors in our model to cancel out.

How does the static perturbation changing the eigenvalues affect this argument? Consider first the effect of a perturbation on the initial ground state and expand its effect using perturbation theory. In particular if $|\psi_g(0)\rangle$ is the initial ground state of the unperturbed system and $|\psi_g^\prime(0)\rangle$ is the initial ground state of the perturbed system, then the first-order correction will be 
\begin{align}
	|\psi_g^\prime(0)\rangle \approx &\ |\psi_g(0)\rangle +\lambda  \sum_{k \neq g} \frac{\langle \psi_k(0)|V|\psi_g(0)\rangle}{E_g -E_k } |\psi_k(0)\rangle \nonumber \\
	& \quad +O(\lambda^2)  \label{eq:perturb}
\end{align}
where $|\psi_k(0)\rangle$ are the excited energy states of the unperturbed system. Now notice that if $V$ is a sum of local terms, each local term will act to change the eigenvalues of at most three vertex operators. Thus, the sum above will not be over all excited states, but will only include excited states with at most three vertex operator eigenvalues flipped. Each of these terms represents errors that are localized in space-time on the quantum circuit version of the quantum wire, as per our argument in the previous subsection. If we were to end up in the $+1$ eigenstate of the final Hamiltonian's $[X]_i$ terms, these terms would then each represent a wire in which at most three single-qubit errors occurred during the evolution. In other words, the portion of the wavefunction arising to first order in perturbation theory, if dragged to the $+1$ eigenstates of the $[X]_i$ operators, acts as if one had (nearly) independently erred the quantum information transmitted through the quantum circuit. A similar argument can be made for the effect of being in the wrong final state. Thus we see that the effect of being in the wrong initial and final eigenspace is, for a weak local perturbation, equivalent to an error that is local in space and time in the quantum circuit. 

Having shown that the effects of changing the eigenvectors of the ground state can be modeled as an independent error model on the quantum wire, we now turn to the effect of static perturbations on the eigenvalues. The worry here is as follows: the effect of the perturbation may cause the energy gap to close and therefore the adiabatic evolution will actually cause the evolution to not preserve the ground state (subspace). This will certainly happen in the system, but the real question is whether this looks like an independent error model or not.  

\begin{figure}[t]
\centering
\includegraphics[width=3.5in]{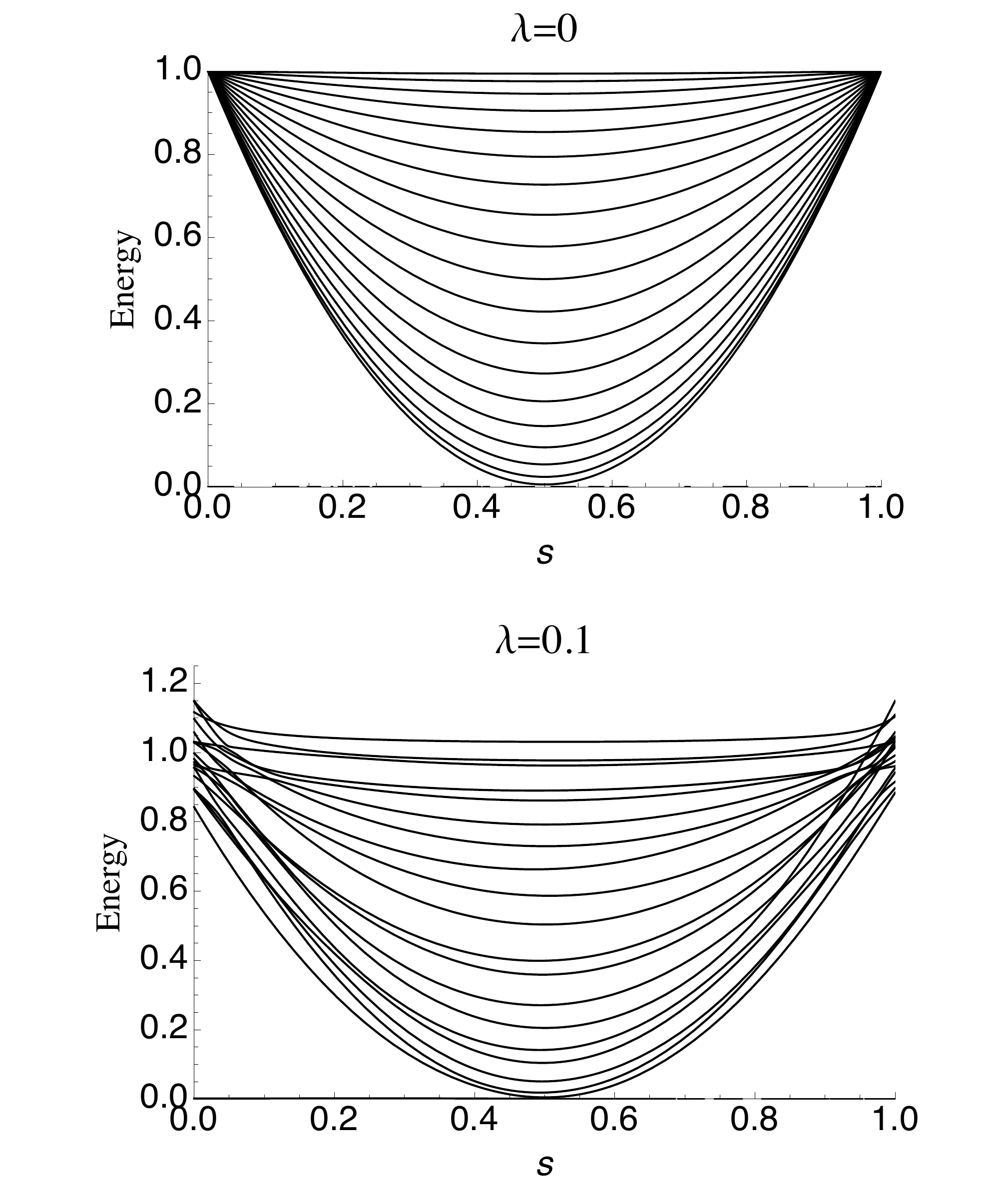}
\caption{Static perturbations. Consider a quantum wire with repeated Hadamard gates, but where the individual interactions in, for example Eq.~(\ref{eq:wire2}) are randomly perturbed by a random amount in the range of $-\lambda$ to $\lambda$. This plot is done for a wire of length $20$. This model can be mapped to the transverse Ising model and above we plot the energy of the free fermions that occur in this model. One sees that while the energy cost for each fermion with the perturbed field is modified, qualitatively the cost for each individual fermion remains approximately the same. This result implies that the model will behave similarly to the unperturbed model: in the absence of other errors the worst this can do is cause a level crossing near $s=1/2$ which will at most introduce a constant number of errors in the model. Furthermore, the effects of other errors occurring on this perturbed model will be similar to how we argue they will act in the main text. In particular, there will be a constant fraction of errors occurring when the system is coupled to a thermal bath and obeys detailed balance.}
\label{fig:pert}
\end{figure}

To see that this is unlikely to be a problem, consider a model in which each of the individual terms in the Hamiltonian of the quantum wire, Eq.~(\ref{eq:wire2}) is randomly perturbed by $\pm \lambda$. When we convert this model into the free-fermion model as discussed in previous sections, we see that this will result in $A$ and $B$ matrices from Eq.~(\ref{eq:AB}) that are like the band diagonal matrices in Eqs.~(\ref{eq:A}) and (\ref{eq:B}), but now with random diagonal and off-diagonal elements added to these matrices. One can then numerically diagonalize the equation for the free fermion energies. In Fig.~\ref{fig:pert}, we show both the unperturbed free-fermion energies and also the perturbed energies for a chain of length $20$ and a perturbation strength of $\lambda=0.1$. From this one sees that while the perturbation causes the energies associated with the free fermions to change by $\pm \lambda$, the spectrum is qualitatively the same. In particular, we can now imagine sweeping adiabatically from $s=0$ to $s=1$. The energy levels that will cause problems are then the free fermions with excitation energies below $\lambda$. At all stages where the fermion energies are greater than $\lambda$ there is a cost to create such a fermion that is greater than $\lambda$, and the perturbing potential $V$ will not be strong enough to achieve this transition. Since only a constant fraction of the energy levels cross below the $\lambda$ creation energy line (similar to Fig.~\ref{fig:figure7}), this means that the possible errors created are a constant fraction of those that could occur on the quantum wire. This, then, is nothing more than an independent error model on the system. Thus, at least for the kind of error model we have assumed, static perturbations result in an independent error model. 

We have argued (but not rigorously proven) in this subsection that the effect of a static perturbation on a quantum wire is no different from an independent error model on the quantum circuit corresponding to this wire. 

\subsubsection{Summary}

In this section we have argued that errors on quantum adiabatic transistors configured as a quantum wire act similarly to errors within the standard quantum circuit model with independent (or slightly time-correlated) errors. Errors that change energy were argued to only act as a constant fraction of independent errors in the quantum circuit model. Errors that preserved the degeneracy could be severe, but only during the initial and final durations of the adiabatic evolution when the quantum information is localized on either end of the quantum wire. By spending less time in the adiabatic evolution during these stages, we showed that these errors act as constant rate errors at the beginning and end of the computation. Finally, we analyzed how static (not varying in time) perturbative interactions would modify these arguments and gave evidence that these errors behave in a similar manner. Thus, adiabatic quantum transistors configured as quantum wires seem to have errors that act as independent single-qubit errors in the standard quantum circuit model. The case of a full quantum circuit with multiple qubits is more difficult to analytically or numerically analyze. Some of our arguments, for instance the effect of local perturbations on the initial twisted cluster-state Hamiltonian that are local in the space-time of the quantum circuit enacted by the adiabatic evolution, carry over directly to this more complicated case.  The full general case is an open question for future work.

\section{Perturbation Theory Gadgets} \label{sec:gadgets}

Finally let us mention issues arising from the use of perturbation theory gadgets in implementing the many-qubit interactions in $H_C$. Perturbation gadgets produce an effective, low-energy many-qubit interaction via strong and weak single- and two-qubit interactions. The strength of these interactions is related to the ratio of the weak and strong interactions (hence the name.) For example, one can engineer an effective three-body interaction at second-order in perturbation theory by using only physically realistic two-body interactions. 

If one wishes to only apply fields (single-qubit interactions) to our model, it appears that one cannot use Bartlett-Rudolph gadgets~\cite{Bartlett:06a}, but instead must use gadgets that do not use encoded quantum information~\cite{Oliveira:08a} (this is because the former gadgets have encoded quantum information that is not acted upon by single-qubit operations.) At temperatures much lower than the energy gap in the ancillas of these constructions, one can effectively treat the errors arising from these gadgets as local errors in our above model. In particular, as in prior work~\cite{Bartlett:06a, Oliveira:08a, Bacon:09a}, we can treat the imperfections arising from the use of perturbation gadgets as terms that cause errors on our system. Thus the only major quantitative change resulting from using these gadgets is that the strengths of the engineered interactions will be lower because of the perturbative nature of these gadgets. The best possible choice of gadget will depend on the physical system being used and on the relevant noise sources that are present.

The main question that arises from the use of perturbation theory gadgets is whether they will cause extra problems when the quantum transistor is configured to carry out a fault-tolerant quantum circuit element. In the supplementary section of Ref.~\cite{Bacon:09a}, we analyzed the effects of perturbation theory gadgets for implementing a simple adiabatic scheme that performs a two-qubit gate. The main result of this analysis was that the use of perturbation theory gadgets to achieve these many-body interactions does \emph{not} substantially damage the usefulness of these gadgets in the adiabatic constructions. However, the use of gadgets has two drawbacks: (1) the eigenvectors are slightly changed because of the inexact nature of the effective many-qubit interaction, and (2) the eigenvalues are also slightly changed and thus the gap is made slightly smaller. For the case analyzed in Ref.~\cite{Bacon:09a}, there were two energy scales, $\lambda$ and $\Delta$. We can choose units where $\Delta=1$. The effective three-qubit interaction in the scheme is of strength $O(\lambda^2)$, and thus the gap in this construction is of this order. The imperfections in the gadget construction led to corrections to the energy of order $O(\lambda^3)$. Furthermore, the eigenvectors of the ground state were corrected in amplitude to order $O(\lambda^2)$. The main conclusion of this calculation was that the effects of the perturbation theory gadgets were small in the perturbation parameter $\lambda$. 

What can we say about the similar calculation for the adiabatic quantum transistor? The first question is how the gadgets will affect the initial and final eigenvectors of the system. Bartlett and Rudolph~\cite{Bartlett:06a} considered exactly this problem for their encoded gadget scheme, and showed that the new eigenvectors behave as if there is an independent error model acting on the state when it is used for measurement-based quantum computing. One can perform exactly the same sort of calculation for the perturbation theory gadgets that use ancillas as mediators and the conclusion is exactly the same: the lowest-order errors created using these gadgets change the state to act as if it has been independently erred. The result is not dissimilar to that of our Eq.~(\ref{eq:perturb}), but now for the more complicated quantum transistor with a quantum circuit. The other question is what effect the gadgets have on the energy gap. We note here that the corrections to the effective interaction for the perturbation theory gadgets can be thought of as extra static Hamiltonian terms that are weaker by the perturbative factor used to create the gadgets, hence these errors are covered by the argument in the previous section.

The use of perturbation theory gadgets in building adiabatic quantum transistors is clearly one of the largest drawbacks in our scheme and one of the reasons why we consider this work as an outline for how an adiabatic quantum transistor could work. Recent work using, for example, more physically realistic AKLT states~\cite{Brennen:08a, Bartlett:10a, Renes:11a, Miyake2010, Else2012}, however, indicates that these gadgets may not be a necessary component of a quantum transistor construction. 

\section{Building Blocks for Adiabatic Quantum Transistor Architectures}\label{sec:blocks}

Finally, let us give details about some building blocks for adiabatic quantum transistors that will be useful for building a larger quantum computer architecture that closely mimics today's modern clocked classical computer architectures. These include constructions for a system that performs measurements and for a system that can be used to take classical information and use it to route quantum information in a device built with an adiabatic quantum transistor. The former is essential for fault-tolerant constructions, which must purge the entropy of quantum errors, and the latter is important for building architectures that are modular and synchronous.

\subsection{Adiabatic measurement amplifier}\label{sec:amp}

\begin{figure}[t]
\centering
\includegraphics[width=3.5in]{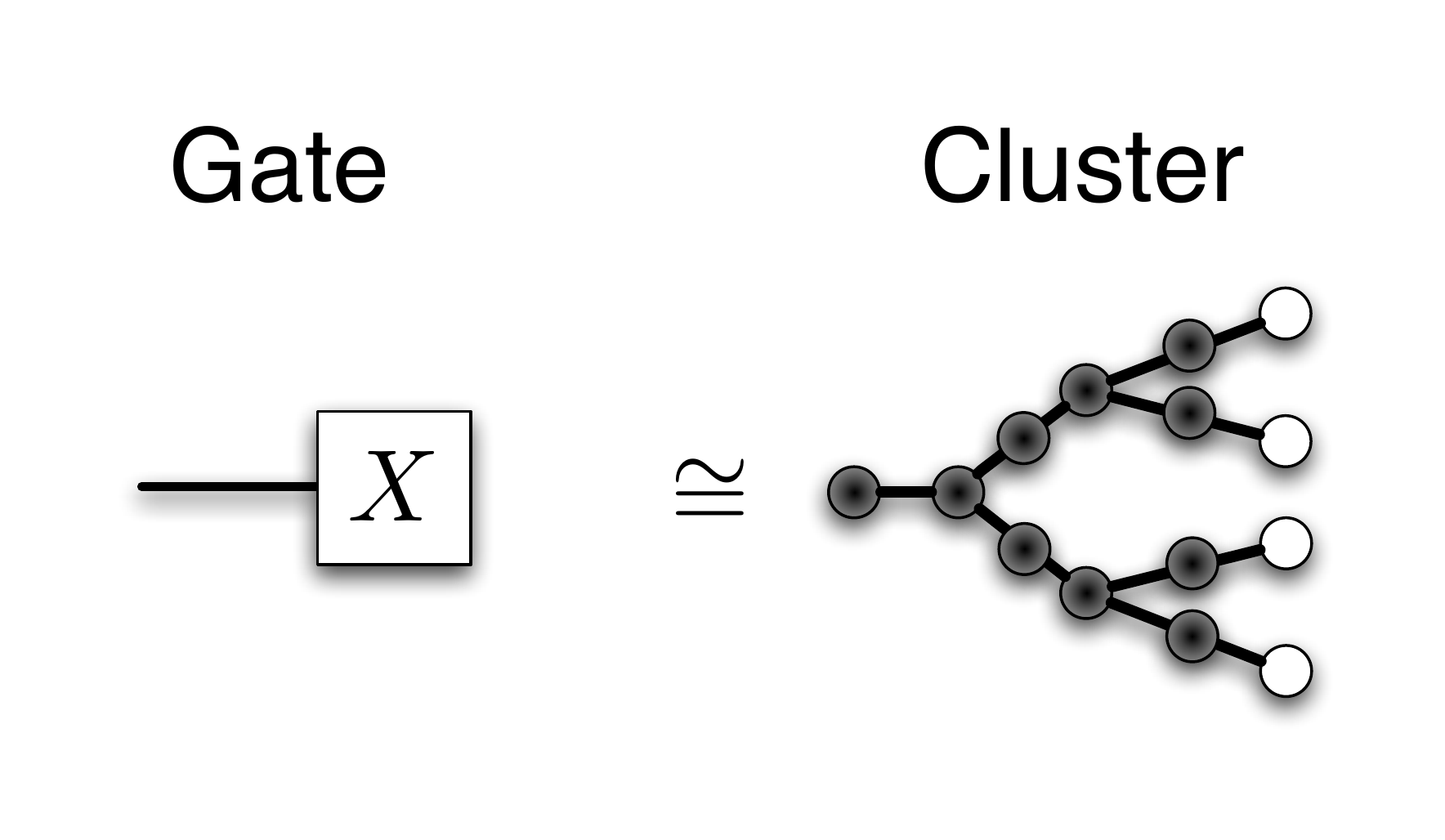}
\caption{An adiabatic measurement amplifier. A binary tree with alternating degree $2$ and $3$ internal vertices (as shown) can be used to take information encoded into the $\bar{X}$ eigenstates at the root into a code across the leaves by turning on an $X$ field across the shaded qubits. This code has a stabilizer given by $Z_i Z_j$ terms on leaves; i.e., it is the repetition code. The logical information initially at the root in the $X$ eigenstate is in this process copied into the $Z$ eigenstates of the leaves. Thus if one turns on a $\sum_i Z_i$ interaction, the information initially in the $X$ eigenstate will be separated in energy by twice the number of leaves. The natural relaxation of this system to its ground state will result in $O( \#~{\rm leaves})$ relaxation events, which can be individually (noisily) observed, hence amplifying a measurement in the $X$ eigenvector basis.}
\label{fig:measure}
\end{figure}

Here we describe the details of an adiabatic measurement amplifier. This amplifier can be used to take quantum information encoded on a single quantum wire (in the circuit being adiabatically simulated) and copy it to multiple qubits in a fixed basis and then read out this information. Consider a tree with alternating internal levels of degree $2$ and $3$ as shown in Fig.~\ref{fig:measure}. Call the degree-1 external node that is connected to an internal degree-3 node the \emph{root} of the tree and label it $r$. Label the unique child node of $r$ by $r'$, and let all other degree-1 nodes be called leaves. Initially, information will have logical operators $\bar{Z}=Z_r$ and $\bar{X}=X_r Z_{r^\prime}$. The operation of the amplifier is as follows. First, single-qubit $X$ terms are turned on while turning off all of the graph stabilizers. Then a $\sum_i Z_i$ term is turned on across all of the leaves. One then waits for observed decay events from the leaves: no decays indicates the system was in the $-1$ eigenstate of $\bar{X}$ initially; $O(\#{\rm leaves})$ decays indicates the system was in the $+1$ eigenstate of $\bar{X}$ initially. The utility of this device is that it turns information encoded into one qubit into information copied (in a particular basis) to many qubits, thus facilitating the measurement of this information.

To see how this process works, we note that, while we defined the information as initially encoded into $\bar{Z}=Z_r$ and $\bar{X}=X_r Z_{r^\prime}$, we could have chosen equivalent operators by multiplying by graph stabilizers. Thus, if we multiply the $\bar{Z}$ operator by graph stabilizers from vertices of degree $3$ and the graph stabilizer from the leaves, we obtain an operator that has $X$ or $I$ on the non-leaf vertices, and is $X$ on all of the leaves. Similarly, if we multiply a $\bar{X}$ operator by a graph stabilizer from vertices of degree $2$ on any simple path from a root to a leaf we will obtain $X$ operators on the non-leaves, and a $Z$ on the leaf. This latter fact is independent of which path to the leaf one chooses, which is a symptom of the fact that the information is encoded into an error-correcting code. Indeed it is not hard to see that by multiplying graph stabilizer elements one can obtain $Z_i Z_j$ terms that act only on $i$ and $j$ which are leaves of the graph. These are the stabilizer elements for the classical repetition code: $|b\rangle \rightarrow |b\rangle^{\otimes k}$. When we turn on the $X$ terms on the non-leaves we will thus have taken the information at the root (and its child) and encoded it into this code (note that the roles of $\bar{X}$ and $\bar{Z}$ are reversed in this process because of an extra Hadamard at the beginning of this process). Stabilizer code arguments~\cite{Bacon:09a, Bacon:09b} make this statement rigorous: we promote the graph stabilizer to logical $Z$s and then the local $X$ terms become products of logical $X$s. The information as described by the encoding above is untouched by this process and thus the adiabatic evolution described above does not affect this encoded information. 

An equivalent way to derive this graph is to take the circuit for copying information recursively using controlled-NOTs and prepared ancilla states, convert it to a graph state, and then simplify this state by noting, for instance, that two nodes on a line can be eliminated from a graph because they correspond to $H^2=I$, where $H$ is the Hadamard gate.

\subsection{Adiabatic Router}

\begin{figure}[t]
\centering
\includegraphics[width=3.5in]{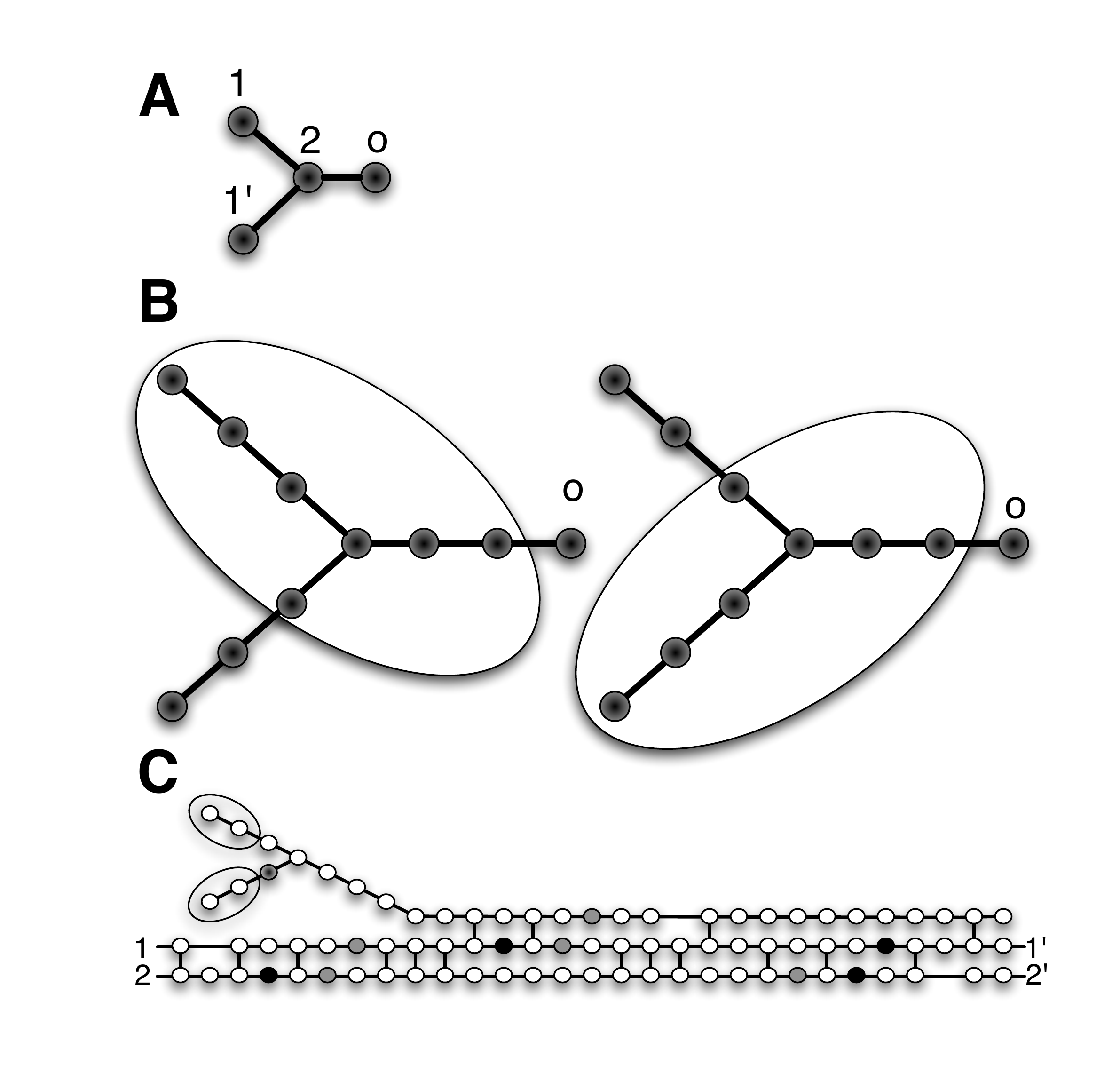}
\caption{Gadgets and Router. Here we show the stages in understanding a router. In {(A)} we show the basic gadget that can be used to conditionally prepare a $\pm 1$ eigenstate of $X_o$, depending on where the field is applied. {(B)} is the same construction, but now demonstrating the flexibility of the construction to the applied field. The applied field need only cover the desired leg in order to get the correct conditional behavior. {(C)} is a non-optimized router construction. If we apply the field everywhere but one of the two circled regions, this will \emph{route} the quantum information incoming at points $1$ and $2$ and output it at points $1'$ and $2'$, applying either the identity gate $(1,2) \to (1',2')$ or the swap gate, $(1,2) \to (2',1')$, depending on which circled region contains a nonzero field. Here the black nodes are twisted by $\frac{\pi}{4}$, the grey nodes by $-\frac{\pi}{4}$, and the gradient-shaded node by $\pi$.}
\label{fig:router}
\end{figure}

It is convenient to design a device that can route quantum information in making a modular clocked architecture. The goal of the router construction is to design a system such that application of a field across different portions of the device can be used to steer the quantum information conditional on \emph{where} the field has been applied. The basic method for achieving this goal is the gadget shown in Fig.~\ref{fig:router}. The gadget shown can be used to produce a state, conditional on where in the device a field is applied. Initially the Hamiltonian for this gadget is a sum over graph stabilizers for all vertices:
\begin{align}
H_i = - (Z_2 X_1 - Z_2 X_{1^\prime} + Z_2 X_o + Z_1 Z_{1^\prime} Z_o X_2)
\end{align}
Then this Hamiltonian is adiabatically turned off while either of the two following Hamiltonians is turned on:
\begin{align}
H_f= - (X_1+X_2)
\end{align}
or
\begin{align}
H_f^\prime =-(X_{1^\prime} + X_2)\,.
\end{align}
The point of this process is that, depending on which of these is turned on, the state of the qubit $o$ will differ between being a $+1$ eigenstate of $X$ and a $-1$ eigenstate of $X$. 

To see this result, first consider the case of ending in $H_f$. Define a stabilizer code using the following operators
\begin{eqnarray}
S_1&=& X_1 X_{1^\prime} , ~ S_2 = X_1 X_o , \nonumber \\
 \bar{Z}_1&=&Z_1 Z_{1^\prime} Z_o X_2, ~\bar{X}_1=X_1,  \nonumber \\  \bar{Z}_2&=&Z_2 X_1,~\bar{X}_2 =X_2 
\end{eqnarray}
where $S_1$ and $S_2$ are the stabilizers for the code and $\bar{P}_j$ are the encoded Pauli operators for this code. Then we see that we can express the initial Hamiltonian as
\begin{equation}
H_i= - (\bar{Z}_1+\bar{Z}_2(I-S_1+S_2))
\end{equation} 
and the final Hamiltonian as
\begin{equation}
H_f = - (\bar{X}_1+ \bar{X}_2)
\end{equation}
Since $S_1$ and $S_2$ commute with these Hamiltonians, the subspaces defined by the eigenvalues of these operators will remain constant. The initial Hamiltonian will have a ground state that is in the $-1$ eigenstate of $S_1$ and the $+1$ eigenstate of $S_2$. Initially the system will also be in the $+1$ eigenvalue eigenspace of $\bar{Z}_1$ and $\bar{Z}_2$. Upon application of the fields as represented by $H_f$, the information in these later two encoded qubits will be adiabatically dragged (with no energy level crossings) to $+1$ eigenvalues of $\bar{X}_1$ and $\bar{X}_2$. Thus at the end of this evolution the system will be in the $+1$ eigenvalue eigenspaces of $S_2$, $\bar{X}_1$, and $\bar{X}_2$, while being in the $-1$ eigenvalue eigenspace of $S_1$. In particular, $X_o=\bar{X}_1 S_2$, which implies that the system is in the eigenstate of $X_o$ with eigenvalue $+1$.

On the contrary, consider the case ending in $H_f^\prime$. Define the code similarly to above,
\begin{eqnarray}
	S_1^\prime &=& X_1 X_{1^\prime} , S_2^\prime = X_{1^\prime} X_o , \nonumber \\
	\bar{Z}_1^\prime&=&Z_1 Z_{1^\prime} Z_o X_2 ,\bar{X}_1^\prime=X_{1^\prime}, \nonumber \\
	\bar{Z}_2^\prime & =& Z_2 X_{1^\prime}, \bar{X}_2 =X_2 \,.
\end{eqnarray}
Then we can express the initial Hamiltonian as
\begin{equation}
	H_i = - \bigl(\bar{Z}_1^\prime+\bar{Z}_2^\prime(-I+S_1^\prime+S_2^\prime)\bigr)
\end{equation}
and the final, primed, Hamiltonian as
\begin{equation}
	H_f^\prime=- \bigl(\bar{X}_1^\prime + \bar{X}_2^\prime\bigr).
\end{equation}
From these expressions one can deduce that the system will start in the $-1$ eigenstate of the $\bar{Z}_2^\prime$, $S_1^\prime$, and $S_2^\prime$ operators and in the $+1$ eigenstate of $\bar{Z}_1^\prime$. At the end of the evolution the system will end up in the $+1$ eigenstate of $\bar{X}_1^\prime$ and $\bar{X}_2^\prime$ and in the $-1$ eigenstate of $S_1^\prime$ and $S_2^\prime$. Since $X_o=\bar{X}_1^\prime S_2^\prime$, this implies that at the end of this adiabatic evolution the system will be in the $-1$ eigenstate of $X_o$. 

Thus we see that depending on whether the final Hamiltonian is $H_f$ or $H_f^\prime$, the qubit located at $o$ is in either the $+1$ or $-1$ eigenstate of $X_o$. Note that this depends only on where the field $H_f$ is applied. Furthermore, note that there is flexibility in spatially achieving this result. For example, in Fig.~\ref{fig:router}(B) we show a larger version of this gadget. Depending on whether the applied field is in either of the two circles, this produces a $+1$ or $-1$ eigenstate of the last qubit at $o$. The exact location of this applied field is not important, except for the point that the field entirely spans one of the two ``legs'' in the construction. Of course, while we would ideally have a field profile that exactly vanishes outside the oval in Fig.~\ref{fig:router}, imperfections in the field will affect the qubits on the boundary; we leave open the question of quantifying the errors introduced by realistic control fields. 

One can extend this idea to then create a router where the routing depends only upon where a field has been applied. The basic idea is rather simple: if one has the ability to conditionally create one of the two orthogonal states by the location where the field is applied, then one can use this as input into a controlled swap gate. In Fig.~\ref{fig:router}(C) we show, for example, such a construction (no attempt has been made to optimize this construction) based upon the conditional swap gate construction of Smolin and DiVincenzo~\cite{Smolin:96a}. If one applies a field everywhere but in one of the two circled regions, then this routes the quantum information depending on which of the two circled regions is left out. In particular, information coming into $1$ and $2$ is thus routed (permuted) to output information at $1^\prime$ and $2^\prime$.

\section{Conclusion} \label{sec:conc}

In conclusion, we have introduced a new method for building a quantum computer based upon the notion of an adiabatic quantum transistor. Two notable benefits of this method are that the system is robust to timing errors (as in universal adiabatic QC) and that it is modular in nature (something that prior universal adiabatic QC models lacked). Instead of requiring increasingly accurate timing and control mechanisms, this model requires one to focus on increasing the fabrication quality of engineered interactions in many-qubit systems. We have argued that the noise model for our scheme will follow an independent error model and thus is amenable to stabilization by standard methods of fault-tolerant quantum computing. While our actual construct is not optimized for current experimental implementation, the mere existence of devices like the one we describe, combined with  recent experimental progress in building highly controllable quantum simulators~\cite{Buluta:09a}, gives us hope that adiabatic quantum transistors are a viable new path toward building a large-scale quantum computer.

\acknowledgments

DB was supported by the NSF under grants 0803478, 0829937, and 0916400. DB and GMC were supported by DARPA under QuEST grant FA-9550-09-1-0044. STF was supported by the Australian Research Council Centre of Excellence for Engineered Quantum Systems CE110001013 and by the Office of the Director of National Intelligence and Intelligence Advanced Research Projects Activity (IARPA) through the Army Research Office. GMC was supported by a Computational Science Graduate Fellowship under DoE grant DE-FG02-97ER25308. 

\bibliography{aqt}

\begin{thebibliography}{46}%
\makeatletter
\providecommand \@ifxundefined [1]{%
 \@ifx{#1\undefined}
}%
\providecommand \@ifnum [1]{%
 \ifnum #1\expandafter \@firstoftwo
 \else \expandafter \@secondoftwo
 \fi
}%
\providecommand \@ifx [1]{%
 \ifx #1\expandafter \@firstoftwo
 \else \expandafter \@secondoftwo
 \fi
}%
\providecommand \natexlab [1]{#1}%
\providecommand \enquote  [1]{``#1''}%
\providecommand \bibnamefont  [1]{#1}%
\providecommand \bibfnamefont [1]{#1}%
\providecommand \citenamefont [1]{#1}%
\providecommand \href@noop [0]{\@secondoftwo}%
\providecommand \href [0]{\begingroup \@sanitize@url \@href}%
\providecommand \@href[1]{\@@startlink{#1}\@@href}%
\providecommand \@@href[1]{\endgroup#1\@@endlink}%
\providecommand \@sanitize@url [0]{\catcode `\\12\catcode `\$12\catcode
  `\&12\catcode `\#12\catcode `\^12\catcode `\_12\catcode `\%12\relax}%
\providecommand \@@startlink[1]{}%
\providecommand \@@endlink[0]{}%
\providecommand \url  [0]{\begingroup\@sanitize@url \@url }%
\providecommand \@url [1]{\endgroup\@href {#1}{\urlprefix }}%
\providecommand \urlprefix  [0]{URL }%
\providecommand \Eprint [0]{\href }%
\providecommand \doibase [0]{http://dx.doi.org/}%
\providecommand \selectlanguage [0]{\@gobble}%
\providecommand \bibinfo  [0]{\@secondoftwo}%
\providecommand \bibfield  [0]{\@secondoftwo}%
\providecommand \translation [1]{[#1]}%
\providecommand \BibitemOpen [0]{}%
\providecommand \bibitemStop [0]{}%
\providecommand \bibitemNoStop [0]{.\EOS\space}%
\providecommand \EOS [0]{\spacefactor3000\relax}%
\providecommand \BibitemShut  [1]{\csname bibitem#1\endcsname}%
\let\auto@bib@innerbib\@empty
\bibitem [{\citenamefont {Shockley}(1948)}]{Shockley:48a}%
  \BibitemOpen
  \bibfield  {author} {\bibinfo {author} {\bibfnamefont {W.}~\bibnamefont
  {Shockley}},\ }\href@noop {} {\emph {\bibinfo {title} {Circuit Element
  Utilizing Semiconductor Material}}},\ \bibinfo {type} {U.S. Patent}\ \bibinfo
  {number} {2569347}\ (\bibinfo  {institution} {Bell Telephone Laboratories},\
  \bibinfo {year} {1948})\BibitemShut {NoStop}%
\bibitem [{\citenamefont {Shor}(1994)}]{Shor:94a}%
  \BibitemOpen
  \bibfield  {author} {\bibinfo {author} {\bibfnamefont {P.~W.}\ \bibnamefont
  {Shor}},\ }\bibfield  {title} {\enquote {\bibinfo {title} {Algorithms for
  quantum computation: Discrete log and factoring},}\ }in\ \href {\doibase
  10.1109/SFCS.1994.365700} {\emph {\bibinfo {booktitle} {Proceedings of the
  35th Annual Symposium on the Foundations of Computer Science}}},\ \bibinfo
  {editor} {edited by\ \bibinfo {editor} {\bibfnamefont {S.}~\bibnamefont
  {Goldwasser}}}\ (\bibinfo  {publisher} {IEEE Computer Society},\ \bibinfo
  {address} {Los Alamitos, CA},\ \bibinfo {year} {1994})\ pp.\ \bibinfo {pages}
  {124--134}\BibitemShut {NoStop}%
\bibitem [{\citenamefont {Aharonov}\ and\ \citenamefont
  {Ben-Or}(1997)}]{Aharonov:97a}%
  \BibitemOpen
  \bibfield  {author} {\bibinfo {author} {\bibfnamefont {D.}~\bibnamefont
  {Aharonov}}\ and\ \bibinfo {author} {\bibfnamefont {M.}~\bibnamefont
  {Ben-Or}},\ }\bibfield  {title} {\enquote {\bibinfo {title} {Fault-tolerant
  quantum computation with constant error rate},}\ }in\ \href {\doibase
  10.1137/S0097539799359385} {\emph {\bibinfo {booktitle} {Proceedings of the
  29th Annual ACM Symposium on Theory of Computing}}}\ (\bibinfo  {publisher}
  {ACM Press},\ \bibinfo {year} {1997})\ pp.\ \bibinfo {pages}
  {176--188}\BibitemShut {NoStop}%
\bibitem [{\citenamefont {Knill}\ \emph
  {et~al.}(1998{\natexlab{a}})\citenamefont {Knill}, \citenamefont {Laflamme},\
  and\ \citenamefont {Zurek}}]{Knill:98a}%
  \BibitemOpen
  \bibfield  {author} {\bibinfo {author} {\bibfnamefont {E.}~\bibnamefont
  {Knill}}, \bibinfo {author} {\bibfnamefont {R.}~\bibnamefont {Laflamme}}, \
  and\ \bibinfo {author} {\bibfnamefont {W.~H.}\ \bibnamefont {Zurek}},\
  }\bibfield  {title} {\enquote {\bibinfo {title} {Resilent quantum
  computation},}\ }\href {\doibase 10.1126/science.279.5349.342} {\bibfield
  {journal} {\bibinfo  {journal} {Science}\ }\textbf {\bibinfo {volume}
  {279}},\ \bibinfo {pages} {342--345} (\bibinfo {year}
  {1998}{\natexlab{a}})}\BibitemShut {NoStop}%
\bibitem [{\citenamefont {Knill}\ \emph
  {et~al.}(1998{\natexlab{b}})\citenamefont {Knill}, \citenamefont {Laflamme},\
  and\ \citenamefont {Zurek}}]{Knill:98b}%
  \BibitemOpen
  \bibfield  {author} {\bibinfo {author} {\bibfnamefont {E.}~\bibnamefont
  {Knill}}, \bibinfo {author} {\bibfnamefont {R.}~\bibnamefont {Laflamme}}, \
  and\ \bibinfo {author} {\bibfnamefont {W.~H.}\ \bibnamefont {Zurek}},\
  }\bibfield  {title} {\enquote {\bibinfo {title} {Resilient quantum
  computation: error models and thresholds},}\ }\href {\doibase
  10.1098/rspa.1998.0166} {\bibfield  {journal} {\bibinfo  {journal} {Proc.
  Roy. Soc. London Ser. A}\ }\textbf {\bibinfo {volume} {454}},\ \bibinfo
  {pages} {365--384} (\bibinfo {year} {1998}{\natexlab{b}})}\BibitemShut
  {NoStop}%
\bibitem [{\citenamefont {Deutsch}(1989)}]{Deutsch:89a}%
  \BibitemOpen
  \bibfield  {author} {\bibinfo {author} {\bibfnamefont {D.}~\bibnamefont
  {Deutsch}},\ }\bibfield  {title} {\enquote {\bibinfo {title} {Quantum
  computational networks},}\ }\href {\doibase 10.1098/rspa.1989.0099}
  {\bibfield  {journal} {\bibinfo  {journal} {Proc. Roy. Soc. London Ser. A}\
  }\textbf {\bibinfo {volume} {425}},\ \bibinfo {pages} {73--90} (\bibinfo
  {year} {1989})}\BibitemShut {NoStop}%
\bibitem [{\citenamefont {Bose}(2003)}]{Bose:03a}%
  \BibitemOpen
  \bibfield  {author} {\bibinfo {author} {\bibfnamefont {Sougato}\ \bibnamefont
  {Bose}},\ }\bibfield  {title} {\enquote {\bibinfo {title} {Quantum
  communication through an unmodulated spin chain},}\ }\href {\doibase
  10.1103/PhysRevLett.91.207901} {\bibfield  {journal} {\bibinfo  {journal}
  {Phys. Rev. Lett.}\ }\textbf {\bibinfo {volume} {91}},\ \bibinfo {pages}
  {207901} (\bibinfo {year} {2003})}\BibitemShut {NoStop}%
\bibitem [{\citenamefont {Knill}\ \emph {et~al.}(2001)\citenamefont {Knill},
  \citenamefont {Laflamme},\ and\ \citenamefont {Milburn}}]{Knill:01a}%
  \BibitemOpen
  \bibfield  {author} {\bibinfo {author} {\bibfnamefont {E.}~\bibnamefont
  {Knill}}, \bibinfo {author} {\bibfnamefont {R.}~\bibnamefont {Laflamme}}, \
  and\ \bibinfo {author} {\bibfnamefont {G.~J.}\ \bibnamefont {Milburn}},\
  }\bibfield  {title} {\enquote {\bibinfo {title} {A scheme for efficient
  quantum computation with linear optics},}\ }\href {\doibase 10.1038/35051009}
  {\bibfield  {journal} {\bibinfo  {journal} {Nature}\ }\textbf {\bibinfo
  {volume} {409}},\ \bibinfo {pages} {46--52} (\bibinfo {year}
  {2001})}\BibitemShut {NoStop}%
\bibitem [{\citenamefont {Raussendorf}\ and\ \citenamefont
  {Briegel}(2001)}]{Raussendorf:01a}%
  \BibitemOpen
  \bibfield  {author} {\bibinfo {author} {\bibfnamefont {R.}~\bibnamefont
  {Raussendorf}}\ and\ \bibinfo {author} {\bibfnamefont {H.~J.}\ \bibnamefont
  {Briegel}},\ }\bibfield  {title} {\enquote {\bibinfo {title} {A one-way
  quantum computer},}\ }\href {\doibase 10.1103/PhysRevLett.86.5188} {\bibfield
   {journal} {\bibinfo  {journal} {Phys. Rev. Lett.}\ }\textbf {\bibinfo
  {volume} {86}},\ \bibinfo {pages} {5188--5191} (\bibinfo {year}
  {2001})}\BibitemShut {NoStop}%
\bibitem [{\citenamefont {Aharonov}\ \emph {et~al.}(2004)\citenamefont
  {Aharonov}, \citenamefont {van Dam}, \citenamefont {Kempe}, \citenamefont
  {Landau}, \citenamefont {Lloyd},\ and\ \citenamefont {Regev}}]{Aharonov:04a}%
  \BibitemOpen
  \bibfield  {author} {\bibinfo {author} {\bibfnamefont {D.}~\bibnamefont
  {Aharonov}}, \bibinfo {author} {\bibfnamefont {W.}~\bibnamefont {van Dam}},
  \bibinfo {author} {\bibfnamefont {J.}~\bibnamefont {Kempe}}, \bibinfo
  {author} {\bibfnamefont {Z.}~\bibnamefont {Landau}}, \bibinfo {author}
  {\bibfnamefont {S.}~\bibnamefont {Lloyd}}, \ and\ \bibinfo {author}
  {\bibfnamefont {O.}~\bibnamefont {Regev}},\ }\bibfield  {title} {\enquote
  {\bibinfo {title} {Adiabatic quantum computation is equivalent to standard
  quantum computation},}\ }in\ \href {\doibase 10.1109/FOCS.2004.8} {\emph
  {\bibinfo {booktitle} {45th Annual IEEE Symposium on Foundations of Computer
  Science}}}\ (\bibinfo  {publisher} {IEEE Computer Society},\ \bibinfo
  {address} {Los Alamitos, CA, USA},\ \bibinfo {year} {2004})\ pp.\ \bibinfo
  {pages} {42--51}\BibitemShut {NoStop}%
\bibitem [{\citenamefont {Kempe}\ \emph {et~al.}(2006)\citenamefont {Kempe},
  \citenamefont {Kitaev},\ and\ \citenamefont {Regev}}]{Kempe:06a}%
  \BibitemOpen
  \bibfield  {author} {\bibinfo {author} {\bibfnamefont {J.}~\bibnamefont
  {Kempe}}, \bibinfo {author} {\bibfnamefont {A.}~\bibnamefont {Kitaev}}, \
  and\ \bibinfo {author} {\bibfnamefont {O.}~\bibnamefont {Regev}},\ }\bibfield
   {title} {\enquote {\bibinfo {title} {The complexity of the local hamiltonian
  problem},}\ }\href {\doibase 10.1137/S0097539704445226} {\bibfield  {journal}
  {\bibinfo  {journal} {SIAM Journal of Computing}\ }\textbf {\bibinfo {volume}
  {35}},\ \bibinfo {pages} {1070--1097} (\bibinfo {year} {2006})}\BibitemShut
  {NoStop}%
\bibitem [{\citenamefont {Mizel}\ \emph {et~al.}(2007)\citenamefont {Mizel},
  \citenamefont {Lidar},\ and\ \citenamefont {Mitchell}}]{Mizel:06a}%
  \BibitemOpen
  \bibfield  {author} {\bibinfo {author} {\bibfnamefont {A.}~\bibnamefont
  {Mizel}}, \bibinfo {author} {\bibfnamefont {D.~A.}\ \bibnamefont {Lidar}}, \
  and\ \bibinfo {author} {\bibfnamefont {M.~W.}\ \bibnamefont {Mitchell}},\
  }\bibfield  {title} {\enquote {\bibinfo {title} {Simple proof of equivalence
  between adiabatic quantum computation and the circuit model},}\ }\href
  {\doibase 10.1103/PhysRevLett.99.070502} {\bibfield  {journal} {\bibinfo
  {journal} {Phys. Rev. Lett.}\ }\textbf {\bibinfo {volume} {99}},\ \bibinfo
  {pages} {070502} (\bibinfo {year} {2007})}\BibitemShut {NoStop}%
\bibitem [{\citenamefont {Kult}\ \emph {et~al.}(2006)\citenamefont {Kult},
  \citenamefont {{\AA}berg},\ and\ \citenamefont {Sj\"{o}qvist}}]{Kult2006}%
  \BibitemOpen
  \bibfield  {author} {\bibinfo {author} {\bibfnamefont {D.}~\bibnamefont
  {Kult}}, \bibinfo {author} {\bibfnamefont {J.}~\bibnamefont {{\AA}berg}}, \
  and\ \bibinfo {author} {\bibfnamefont {E.}~\bibnamefont {Sj\"{o}qvist}},\
  }\bibfield  {title} {\enquote {\bibinfo {title} {Noncyclic geometric changes
  of quantum states},}\ }\href {\doibase 10.1103/PhysRevA.74.022106} {\bibfield
   {journal} {\bibinfo  {journal} {Phys. Rev. A}\ }\textbf {\bibinfo {volume}
  {74}},\ \bibinfo {pages} {022106} (\bibinfo {year} {2006})}\BibitemShut
  {NoStop}%
\bibitem [{\citenamefont {Wootters}\ and\ \citenamefont
  {Zurek}(1982)}]{Wootters:82a}%
  \BibitemOpen
  \bibfield  {author} {\bibinfo {author} {\bibfnamefont {W.}~\bibnamefont
  {Wootters}}\ and\ \bibinfo {author} {\bibfnamefont {W.}~\bibnamefont
  {Zurek}},\ }\bibfield  {title} {\enquote {\bibinfo {title} {A single quantum
  cannot be cloned},}\ }\href {\doibase 10.1038/299802a0} {\bibfield  {journal}
  {\bibinfo  {journal} {Nature}\ }\textbf {\bibinfo {volume} {299}},\ \bibinfo
  {pages} {802--803} (\bibinfo {year} {1982})}\BibitemShut {NoStop}%
\bibitem [{\citenamefont {Bacon}\ and\ \citenamefont
  {Flammia}(2009)}]{Bacon:09a}%
  \BibitemOpen
  \bibfield  {author} {\bibinfo {author} {\bibfnamefont {D.}~\bibnamefont
  {Bacon}}\ and\ \bibinfo {author} {\bibfnamefont {S.~T.}\ \bibnamefont
  {Flammia}},\ }\bibfield  {title} {\enquote {\bibinfo {title} {Adiabatic gate
  teleportation},}\ }\href {\doibase 10.1103/PhysRevLett.103.120504} {\bibfield
   {journal} {\bibinfo  {journal} {Phys. Rev. Lett.}\ }\textbf {\bibinfo
  {volume} {103}},\ \bibinfo {pages} {120504} (\bibinfo {year}
  {2009})}\BibitemShut {NoStop}%
\bibitem [{\citenamefont {Bacon}\ and\ \citenamefont
  {Flammia}(2010)}]{Bacon:09b}%
  \BibitemOpen
  \bibfield  {author} {\bibinfo {author} {\bibfnamefont {D.}~\bibnamefont
  {Bacon}}\ and\ \bibinfo {author} {\bibfnamefont {S.~T.}\ \bibnamefont
  {Flammia}},\ }\bibfield  {title} {\enquote {\bibinfo {title} {Adiabatic
  cluster state quantum computing},}\ }\href {\doibase
  10.1103/PhysRevA.82.030303} {\bibfield  {journal} {\bibinfo  {journal} {Phys.
  Rev. A}\ }\textbf {\bibinfo {volume} {82}},\ \bibinfo {pages} {030303(R)}
  (\bibinfo {year} {2010})}\BibitemShut {NoStop}%
\bibitem [{\citenamefont {Oreshkov}\ \emph {et~al.}(2009)\citenamefont
  {Oreshkov}, \citenamefont {Brun},\ and\ \citenamefont
  {Lidar}}]{Oreshkov:09a}%
  \BibitemOpen
  \bibfield  {author} {\bibinfo {author} {\bibfnamefont {O.}~\bibnamefont
  {Oreshkov}}, \bibinfo {author} {\bibfnamefont {T.~A.}\ \bibnamefont {Brun}},
  \ and\ \bibinfo {author} {\bibfnamefont {D.~A.}\ \bibnamefont {Lidar}},\
  }\bibfield  {title} {\enquote {\bibinfo {title} {Fault-tolerant holonomic
  quantum computation},}\ }\href {\doibase 10.1103/PhysRevA.53.2855} {\bibfield
   {journal} {\bibinfo  {journal} {Phys. Rev. Lett.}\ }\textbf {\bibinfo
  {volume} {102}},\ \bibinfo {pages} {070502} (\bibinfo {year}
  {2009})}\BibitemShut {NoStop}%
\bibitem [{\citenamefont {Oreshkov}(2009)}]{Oreshkov:09b}%
  \BibitemOpen
  \bibfield  {author} {\bibinfo {author} {\bibfnamefont {O.}~\bibnamefont
  {Oreshkov}},\ }\bibfield  {title} {\enquote {\bibinfo {title} {Holonomic
  quantum computation in subsystems},}\ }\href {\doibase
  10.1103/PhysRevLett.103.090502} {\bibfield  {journal} {\bibinfo  {journal}
  {Phys. Rev. Lett.}\ }\textbf {\bibinfo {volume} {103}},\ \bibinfo {eid}
  {090502} (\bibinfo {year} {2009})}\BibitemShut {NoStop}%
\bibitem [{\citenamefont {Briegel}\ and\ \citenamefont
  {Raussendorf}(2001)}]{Briegel:01a}%
  \BibitemOpen
  \bibfield  {author} {\bibinfo {author} {\bibfnamefont {Hans~J.}\ \bibnamefont
  {Briegel}}\ and\ \bibinfo {author} {\bibfnamefont {Robert}\ \bibnamefont
  {Raussendorf}},\ }\bibfield  {title} {\enquote {\bibinfo {title} {Persistent
  entanglement in arrays of interacting particles},}\ }\href {\doibase
  10.1103/PhysRevLett.86.910} {\bibfield  {journal} {\bibinfo  {journal} {Phys.
  Rev. Lett.}\ }\textbf {\bibinfo {volume} {86}},\ \bibinfo {pages} {910--913}
  (\bibinfo {year} {2001})}\BibitemShut {NoStop}%
\bibitem [{\citenamefont {Bartlett}\ and\ \citenamefont
  {Rudolph}(2006)}]{Bartlett:06a}%
  \BibitemOpen
  \bibfield  {author} {\bibinfo {author} {\bibfnamefont {S.~D.}\ \bibnamefont
  {Bartlett}}\ and\ \bibinfo {author} {\bibfnamefont {T.}~\bibnamefont
  {Rudolph}},\ }\bibfield  {title} {\enquote {\bibinfo {title} {Simple
  nearest-neighbor two-body hamiltonian system for which the ground state is a
  universal resource for quantum computation},}\ }\href {\doibase
  10.1103/PhysRevA.74.040302} {\bibfield  {journal} {\bibinfo  {journal} {Phys.
  Rev. A}\ }\textbf {\bibinfo {volume} {74}},\ \bibinfo {pages} {040302}
  (\bibinfo {year} {2006})}\BibitemShut {NoStop}%
\bibitem [{\citenamefont {Oliveira}\ and\ \citenamefont
  {Terhal}(2008)}]{Oliveira:08a}%
  \BibitemOpen
  \bibfield  {author} {\bibinfo {author} {\bibfnamefont {R.}~\bibnamefont
  {Oliveira}}\ and\ \bibinfo {author} {\bibfnamefont {B.M.}\ \bibnamefont
  {Terhal}},\ }\bibfield  {title} {\enquote {\bibinfo {title} {The complexity
  of quantum spin systems on a two-dimensional square lattice},}\ }\href@noop
  {} {\bibfield  {journal} {\bibinfo  {journal} {Quant. Inform. Comp.}\
  }\textbf {\bibinfo {volume} {8}},\ \bibinfo {pages} {0900} (\bibinfo {year}
  {2008})}\BibitemShut {NoStop}%
\bibitem [{\citenamefont {Schaller}\ \emph {et~al.}(2006)\citenamefont
  {Schaller}, \citenamefont {Mostame},\ and\ \citenamefont
  {Sch\"{u}tzhold}}]{Schaller:06a}%
  \BibitemOpen
  \bibfield  {author} {\bibinfo {author} {\bibfnamefont {Gernot}\ \bibnamefont
  {Schaller}}, \bibinfo {author} {\bibfnamefont {Sarah}\ \bibnamefont
  {Mostame}}, \ and\ \bibinfo {author} {\bibfnamefont {Ralf}\ \bibnamefont
  {Sch\"{u}tzhold}},\ }\bibfield  {title} {\enquote {\bibinfo {title} {General
  error estimate for adiabatic quantum computing},}\ }\href {\doibase
  10.1103/PhysRevA.73.062307} {\bibfield  {journal} {\bibinfo  {journal} {Phys.
  Rev. A}\ }\textbf {\bibinfo {volume} {73}},\ \bibinfo {eid} {062307}
  (\bibinfo {year} {2006})}\BibitemShut {NoStop}%
\bibitem [{\citenamefont {Doherty}\ and\ \citenamefont
  {Bartlett}(2009)}]{Doherty:09a}%
  \BibitemOpen
  \bibfield  {author} {\bibinfo {author} {\bibfnamefont {A.C.}\ \bibnamefont
  {Doherty}}\ and\ \bibinfo {author} {\bibfnamefont {S.D.}\ \bibnamefont
  {Bartlett}},\ }\bibfield  {title} {\enquote {\bibinfo {title} {Identifying
  phases of quantum many-body systems that are universal for quantum
  computation},}\ }\href {\doibase 10.1103/PhysRevLett.103.020506} {\bibfield
  {journal} {\bibinfo  {journal} {Phys. Rev. Lett.}\ }\textbf {\bibinfo
  {volume} {103}},\ \bibinfo {pages} {020506} (\bibinfo {year}
  {2009})}\BibitemShut {NoStop}%
\bibitem [{\citenamefont {Pachos}\ and\ \citenamefont
  {Plenio}(2004)}]{Pachos:04a}%
  \BibitemOpen
  \bibfield  {author} {\bibinfo {author} {\bibfnamefont {J.~K.}\ \bibnamefont
  {Pachos}}\ and\ \bibinfo {author} {\bibfnamefont {M.~B.}\ \bibnamefont
  {Plenio}},\ }\bibfield  {title} {\enquote {\bibinfo {title} {Three-spin
  interactions in optical lattices and criticality in cluster hamiltonians},}\
  }\href {\doibase 10.1103/PhysRevLett.93.056402} {\bibfield  {journal}
  {\bibinfo  {journal} {Phys. Rev. Lett.}\ }\textbf {\bibinfo {volume} {93}},\
  \bibinfo {pages} {056402} (\bibinfo {year} {2004})}\BibitemShut {NoStop}%
\bibitem [{\citenamefont {Lieb}\ \emph {et~al.}(1961)\citenamefont {Lieb},
  \citenamefont {Schultz},\ and\ \citenamefont {Mattis}}]{Lieb:61a}%
  \BibitemOpen
  \bibfield  {author} {\bibinfo {author} {\bibfnamefont {E.}~\bibnamefont
  {Lieb}}, \bibinfo {author} {\bibfnamefont {T.}~\bibnamefont {Schultz}}, \
  and\ \bibinfo {author} {\bibfnamefont {D.}~\bibnamefont {Mattis}},\
  }\bibfield  {title} {\enquote {\bibinfo {title} {Two soluble models of an
  antiferromagnetic chain},}\ }\href {\doibase 10.1016/0003-4916(61)90115-4}
  {\bibfield  {journal} {\bibinfo  {journal} {Ann. Phys.}\ }\textbf {\bibinfo
  {volume} {16}},\ \bibinfo {pages} {407} (\bibinfo {year} {1961})}\BibitemShut
  {NoStop}%
\bibitem [{\citenamefont {Verstraete}\ \emph {et~al.}(2004)\citenamefont
  {Verstraete}, \citenamefont {Porras},\ and\ \citenamefont
  {Cirac}}]{Verstraete:04a}%
  \BibitemOpen
  \bibfield  {author} {\bibinfo {author} {\bibfnamefont {F.}~\bibnamefont
  {Verstraete}}, \bibinfo {author} {\bibfnamefont {D.}~\bibnamefont {Porras}},
  \ and\ \bibinfo {author} {\bibfnamefont {J.~I.}\ \bibnamefont {Cirac}},\
  }\bibfield  {title} {\enquote {\bibinfo {title} {Density matrix
  renormalization group and periodic boundary conditions: A quantum information
  perspective},}\ }\href {\doibase 10.1103/PhysRevLett.93.227205} {\bibfield
  {journal} {\bibinfo  {journal} {Phys. Rev. Lett.}\ }\textbf {\bibinfo
  {volume} {93}},\ \bibinfo {pages} {227205} (\bibinfo {year}
  {2004})}\BibitemShut {NoStop}%
\bibitem [{\citenamefont {White}(1992)}]{White:92a}%
  \BibitemOpen
  \bibfield  {author} {\bibinfo {author} {\bibfnamefont {Steven~R.}\
  \bibnamefont {White}},\ }\bibfield  {title} {\enquote {\bibinfo {title}
  {Density matrix formulation for quantum renormalization groups},}\ }\href
  {\doibase 10.1103/PhysRevLett.69.2863} {\bibfield  {journal} {\bibinfo
  {journal} {Phys. Rev. Lett.}\ }\textbf {\bibinfo {volume} {69}},\ \bibinfo
  {pages} {2863--2866} (\bibinfo {year} {1992})}\BibitemShut {NoStop}%
\bibitem [{\citenamefont {Schollw\"ock}(2005)}]{Schollock:05a}%
  \BibitemOpen
  \bibfield  {author} {\bibinfo {author} {\bibfnamefont {U.}~\bibnamefont
  {Schollw\"ock}},\ }\bibfield  {title} {\enquote {\bibinfo {title} {The
  density-matrix renormalization group},}\ }\href {\doibase
  10.1103/RevModPhys.77.259} {\bibfield  {journal} {\bibinfo  {journal} {Rev.
  Mod. Phys.}\ }\textbf {\bibinfo {volume} {77}},\ \bibinfo {pages} {259--315}
  (\bibinfo {year} {2005})}\BibitemShut {NoStop}%
\bibitem [{\citenamefont {Dorier}\ \emph {et~al.}(2005)\citenamefont {Dorier},
  \citenamefont {Becca},\ and\ \citenamefont {Mila}}]{Dorier:05a}%
  \BibitemOpen
  \bibfield  {author} {\bibinfo {author} {\bibfnamefont {J.}~\bibnamefont
  {Dorier}}, \bibinfo {author} {\bibfnamefont {F.}~\bibnamefont {Becca}}, \
  and\ \bibinfo {author} {\bibfnamefont {F.}~\bibnamefont {Mila}},\ }\bibfield
  {title} {\enquote {\bibinfo {title} {Quantum compass model on the square
  lattice},}\ }\href {\doibase 10.1103/PhysRevB.72.024448} {\bibfield
  {journal} {\bibinfo  {journal} {Phys. Rev. B}\ }\textbf {\bibinfo {volume}
  {72}},\ \bibinfo {pages} {024448} (\bibinfo {year} {2005})}\BibitemShut
  {NoStop}%
\bibitem [{\citenamefont {Paz-Silva}\ \emph {et~al.}(2010)\citenamefont
  {Paz-Silva}, \citenamefont {Brennen},\ and\ \citenamefont
  {Twamley}}]{Paz-Silva:10a}%
  \BibitemOpen
  \bibfield  {author} {\bibinfo {author} {\bibfnamefont {Gerardo~A.}\
  \bibnamefont {Paz-Silva}}, \bibinfo {author} {\bibfnamefont {Gavin~K.}\
  \bibnamefont {Brennen}}, \ and\ \bibinfo {author} {\bibfnamefont {Jason}\
  \bibnamefont {Twamley}},\ }\bibfield  {title} {\enquote {\bibinfo {title}
  {Fault tolerance with noisy and slow measurements and preparation},}\ }\href
  {\doibase 10.1103/PhysRevLett.105.100501} {\bibfield  {journal} {\bibinfo
  {journal} {Phys. Rev. Lett.}\ }\textbf {\bibinfo {volume} {105}},\ \bibinfo
  {pages} {100501} (\bibinfo {year} {2010})}\BibitemShut {NoStop}%
\bibitem [{\citenamefont {Childs}\ \emph {et~al.}(2001)\citenamefont {Childs},
  \citenamefont {Farhi},\ and\ \citenamefont {Preskill}}]{Childs:01a}%
  \BibitemOpen
  \bibfield  {author} {\bibinfo {author} {\bibfnamefont {A.}~\bibnamefont
  {Childs}}, \bibinfo {author} {\bibfnamefont {E.}~\bibnamefont {Farhi}}, \
  and\ \bibinfo {author} {\bibfnamefont {J.}~\bibnamefont {Preskill}},\
  }\bibfield  {title} {\enquote {\bibinfo {title} {Robustness of adiabatic
  quantum computation},}\ }\href {\doibase 10.1103/PhysRevA.65.012322}
  {\bibfield  {journal} {\bibinfo  {journal} {Phys. Rev. A}\ }\textbf {\bibinfo
  {volume} {65}},\ \bibinfo {pages} {012322} (\bibinfo {year}
  {2001})}\BibitemShut {NoStop}%
\bibitem [{\citenamefont {Pfeuty}(1970)}]{Pfeuty:70a}%
  \BibitemOpen
  \bibfield  {author} {\bibinfo {author} {\bibfnamefont {P.}~\bibnamefont
  {Pfeuty}},\ }\bibfield  {title} {\enquote {\bibinfo {title} {The
  one-dimensional ising model with a transverse field},}\ }\href {\doibase
  10.1016/0003-4916(70)90270-8} {\bibfield  {journal} {\bibinfo  {journal}
  {Ann. Phys.}\ }\textbf {\bibinfo {volume} {57}},\ \bibinfo {pages} {79--90}
  (\bibinfo {year} {1970})}\BibitemShut {NoStop}%
\bibitem [{\citenamefont {Roland}\ and\ \citenamefont
  {Cerf}(2002)}]{Roland:02a}%
  \BibitemOpen
  \bibfield  {author} {\bibinfo {author} {\bibfnamefont {J\'er\'emie}\
  \bibnamefont {Roland}}\ and\ \bibinfo {author} {\bibfnamefont {Nicolas~J.}\
  \bibnamefont {Cerf}},\ }\bibfield  {title} {\enquote {\bibinfo {title}
  {Quantum search by local adiabatic evolution},}\ }\href {\doibase
  10.1103/PhysRevA.65.042308} {\bibfield  {journal} {\bibinfo  {journal} {Phys.
  Rev. A}\ }\textbf {\bibinfo {volume} {65}},\ \bibinfo {pages} {042308}
  (\bibinfo {year} {2002})}\BibitemShut {NoStop}%
\bibitem [{\citenamefont {Son}\ \emph {et~al.}(2012)\citenamefont {Son},
  \citenamefont {Amico},\ and\ \citenamefont {Vedral}}]{Son2011}%
  \BibitemOpen
  \bibfield  {author} {\bibinfo {author} {\bibfnamefont {W.}~\bibnamefont
  {Son}}, \bibinfo {author} {\bibfnamefont {L.}~\bibnamefont {Amico}}, \ and\
  \bibinfo {author} {\bibfnamefont {V.}~\bibnamefont {Vedral}},\ }\bibfield
  {title} {\enquote {\bibinfo {title} {Topological order in 1d cluster state
  protected by symmetry},}\ }\href {\doibase 10.1007/s11128-011-0346-7}
  {\bibfield  {journal} {\bibinfo  {journal} {Quant. Inform. Proc.}\ }\textbf
  {\bibinfo {volume} {11}},\ \bibinfo {pages} {1961--1968} (\bibinfo {year}
  {2012})}\BibitemShut {NoStop}%
\bibitem [{\citenamefont {Schuch}\ \emph {et~al.}(2011)\citenamefont {Schuch},
  \citenamefont {P\'{e}rez-Garc\'{\i}a},\ and\ \citenamefont
  {Cirac}}]{Schuch2011}%
  \BibitemOpen
  \bibfield  {author} {\bibinfo {author} {\bibfnamefont {Norbert}\ \bibnamefont
  {Schuch}}, \bibinfo {author} {\bibfnamefont {David}\ \bibnamefont
  {P\'{e}rez-Garc\'{\i}a}}, \ and\ \bibinfo {author} {\bibfnamefont {Ignacio}\
  \bibnamefont {Cirac}},\ }\bibfield  {title} {\enquote {\bibinfo {title}
  {Classifying quantum phases using matrix product states and projected
  entangled pair states},}\ }\href {\doibase 10.1103/PhysRevB.84.165139}
  {\bibfield  {journal} {\bibinfo  {journal} {Phys. Rev. B}\ }\textbf {\bibinfo
  {volume} {84}},\ \bibinfo {pages} {165139} (\bibinfo {year}
  {2011})}\BibitemShut {NoStop}%
\bibitem [{\citenamefont {Chen}\ \emph {et~al.}(2013)\citenamefont {Chen},
  \citenamefont {Gu}, \citenamefont {Liu},\ and\ \citenamefont
  {Wen}}]{Chen2013}%
  \BibitemOpen
  \bibfield  {author} {\bibinfo {author} {\bibfnamefont {Xie}\ \bibnamefont
  {Chen}}, \bibinfo {author} {\bibfnamefont {Zheng-Cheng}\ \bibnamefont {Gu}},
  \bibinfo {author} {\bibfnamefont {Zheng-Xin}\ \bibnamefont {Liu}}, \ and\
  \bibinfo {author} {\bibfnamefont {Xiao-Gang}\ \bibnamefont {Wen}},\
  }\bibfield  {title} {\enquote {\bibinfo {title} {Symmetry protected
  topological orders and the group cohomology of their symmetry group},}\
  }\href {\doibase 10.1103/PhysRevB.87.155114} {\bibfield  {journal} {\bibinfo
  {journal} {Phys. Rev. B}\ }\textbf {\bibinfo {volume} {87}},\ \bibinfo
  {pages} {155114} (\bibinfo {year} {2013})},\ \Eprint
  {http://arxiv.org/abs/1106.4772} {arXiv:1106.4772 [cond-mat.str-el]}
  \BibitemShut {NoStop}%
\bibitem [{\citenamefont {Bravyi}\ \emph {et~al.}(2010)\citenamefont {Bravyi},
  \citenamefont {Hastings},\ and\ \citenamefont {Michalakis}}]{Bravyi2010}%
  \BibitemOpen
  \bibfield  {author} {\bibinfo {author} {\bibfnamefont {Sergey}\ \bibnamefont
  {Bravyi}}, \bibinfo {author} {\bibfnamefont {Matthew}\ \bibnamefont
  {Hastings}}, \ and\ \bibinfo {author} {\bibfnamefont {Spyridon}\ \bibnamefont
  {Michalakis}},\ }\bibfield  {title} {\enquote {\bibinfo {title} {Topological
  quantum order: stability under local perturbations},}\ }\href {\doibase
  10.1063/1.3490195} {\bibfield  {journal} {\bibinfo  {journal} {J. Math.
  Phys.}\ }\textbf {\bibinfo {volume} {51}},\ \bibinfo {pages} {093512}
  (\bibinfo {year} {2010})},\ \Eprint {http://arxiv.org/abs/1001.0344}
  {arXiv:1001.0344 [quant-ph]} \BibitemShut {NoStop}%
\bibitem [{\citenamefont {Michalakis}\ and\ \citenamefont
  {Pytel}(2011)}]{Michalakis2011}%
  \BibitemOpen
  \bibfield  {author} {\bibinfo {author} {\bibfnamefont {S.}~\bibnamefont
  {Michalakis}}\ and\ \bibinfo {author} {\bibfnamefont {J.}~\bibnamefont
  {Pytel}},\ }\bibfield  {title} {\enquote {\bibinfo {title} {Stability of
  frustration-free hamiltonians},}\ }\href@noop {} {\  (\bibinfo {year}
  {2011})},\ \Eprint {http://arxiv.org/abs/1109.1588} {arXiv:1109.1588
  [quant-ph]} \BibitemShut {NoStop}%
\bibitem [{\citenamefont {Else}\ \emph {et~al.}(2012)\citenamefont {Else},
  \citenamefont {Schwarz}, \citenamefont {Bartlett},\ and\ \citenamefont
  {Doherty}}]{Else2012}%
  \BibitemOpen
  \bibfield  {author} {\bibinfo {author} {\bibfnamefont {Dominic~V.}\
  \bibnamefont {Else}}, \bibinfo {author} {\bibfnamefont {Ilai}\ \bibnamefont
  {Schwarz}}, \bibinfo {author} {\bibfnamefont {Stephen~D.}\ \bibnamefont
  {Bartlett}}, \ and\ \bibinfo {author} {\bibfnamefont {Andrew~C.}\
  \bibnamefont {Doherty}},\ }\bibfield  {title} {\enquote {\bibinfo {title}
  {Symmetry-protected phases for measurement-based quantum computation},}\
  }\href {\doibase 10.1103/PhysRevLett.108.240505} {\bibfield  {journal}
  {\bibinfo  {journal} {Phys. Rev. Lett.}\ }\textbf {\bibinfo {volume} {108}},\
  \bibinfo {pages} {240505} (\bibinfo {year} {2012})}\BibitemShut {NoStop}%
\bibitem [{\citenamefont {Miyake}(2010)}]{Miyake2010}%
  \BibitemOpen
  \bibfield  {author} {\bibinfo {author} {\bibfnamefont {Akimasa}\ \bibnamefont
  {Miyake}},\ }\bibfield  {title} {\enquote {\bibinfo {title} {Quantum
  computation on the edge of a symmetry-protected topological order},}\ }\href
  {\doibase 10.1103/PhysRevLett.105.040501} {\bibfield  {journal} {\bibinfo
  {journal} {Phys. Rev. Lett.}\ }\textbf {\bibinfo {volume} {105}},\ \bibinfo
  {pages} {040501} (\bibinfo {year} {2010})}\BibitemShut {NoStop}%
\bibitem [{\citenamefont {Brennen}\ and\ \citenamefont
  {Miyake}(2008)}]{Brennen:08a}%
  \BibitemOpen
  \bibfield  {author} {\bibinfo {author} {\bibfnamefont {G.}~\bibnamefont
  {Brennen}}\ and\ \bibinfo {author} {\bibfnamefont {A.}~\bibnamefont
  {Miyake}},\ }\bibfield  {title} {\enquote {\bibinfo {title}
  {Measurement-based quantum computer in the gapped ground state of a two-body
  hamiltonian},}\ }\href {\doibase 10.1103/PhysRevLett.101.010502} {\bibfield
  {journal} {\bibinfo  {journal} {Phys. Rev. Lett.}\ }\textbf {\bibinfo
  {volume} {101}},\ \bibinfo {pages} {010502} (\bibinfo {year}
  {2008})}\BibitemShut {NoStop}%
\bibitem [{\citenamefont {Bartlett}\ \emph {et~al.}(2010)\citenamefont
  {Bartlett}, \citenamefont {Brennen}, \citenamefont {Miyake},\ and\
  \citenamefont {Renes}}]{Bartlett:10a}%
  \BibitemOpen
  \bibfield  {author} {\bibinfo {author} {\bibfnamefont {S.~D.}\ \bibnamefont
  {Bartlett}}, \bibinfo {author} {\bibfnamefont {G.~K.}\ \bibnamefont
  {Brennen}}, \bibinfo {author} {\bibfnamefont {A.}~\bibnamefont {Miyake}}, \
  and\ \bibinfo {author} {\bibfnamefont {J.}~\bibnamefont {Renes}},\ }\bibfield
   {title} {\enquote {\bibinfo {title} {Quantum computational renormalization
  in the haldane phase},}\ }\href {\doibase 10.1103/PhysRevLett.105.110502}
  {\bibfield  {journal} {\bibinfo  {journal} {Phys. Rev. Lett.}\ }\textbf
  {\bibinfo {volume} {105}},\ \bibinfo {pages} {110502} (\bibinfo {year}
  {2010})}\BibitemShut {NoStop}%
\bibitem [{\citenamefont {Renes}\ \emph {et~al.}(2013)\citenamefont {Renes},
  \citenamefont {Miyake}, \citenamefont {Brennen},\ and\ \citenamefont
  {Bartlett}}]{Renes:11a}%
  \BibitemOpen
  \bibfield  {author} {\bibinfo {author} {\bibfnamefont {Joseph~M.}\
  \bibnamefont {Renes}}, \bibinfo {author} {\bibfnamefont {Akimasa}\
  \bibnamefont {Miyake}}, \bibinfo {author} {\bibfnamefont {Gavin~K.}\
  \bibnamefont {Brennen}}, \ and\ \bibinfo {author} {\bibfnamefont
  {Stephen~D.}\ \bibnamefont {Bartlett}},\ }\bibfield  {title} {\enquote
  {\bibinfo {title} {Holonomic quantum computing in ground states of spin
  chains with symmetry-protected topological order},}\ }\href {\doibase
  10.1088/1367-2630/15/2/025020} {\bibfield  {journal} {\bibinfo  {journal}
  {New J. Phys.}\ }\textbf {\bibinfo {volume} {15}},\ \bibinfo {pages} {025020}
  (\bibinfo {year} {2013})},\ \Eprint {http://arxiv.org/abs/1103.5076}
  {arXiv:1103.5076 [quant-ph]} \BibitemShut {NoStop}%
\bibitem [{\citenamefont {Smolin}\ and\ \citenamefont
  {DiVincenzo}(1996)}]{Smolin:96a}%
  \BibitemOpen
  \bibfield  {author} {\bibinfo {author} {\bibfnamefont {J.~A.}\ \bibnamefont
  {Smolin}}\ and\ \bibinfo {author} {\bibfnamefont {D.~P.}\ \bibnamefont
  {DiVincenzo}},\ }\bibfield  {title} {\enquote {\bibinfo {title} {Five two-bit
  quantum gates are sufficient to implement the quantum fredkin gate},}\ }\href
  {\doibase 10.1103/PhysRevA.53.2855} {\bibfield  {journal} {\bibinfo
  {journal} {Phys. Rev. A}\ }\textbf {\bibinfo {volume} {53}},\ \bibinfo
  {pages} {2855--2856} (\bibinfo {year} {1996})}\BibitemShut {NoStop}%
\bibitem [{\citenamefont {Buluta}\ and\ \citenamefont
  {Nori}(2009)}]{Buluta:09a}%
  \BibitemOpen
  \bibfield  {author} {\bibinfo {author} {\bibfnamefont {Iulia}\ \bibnamefont
  {Buluta}}\ and\ \bibinfo {author} {\bibfnamefont {Franco}\ \bibnamefont
  {Nori}},\ }\bibfield  {title} {\enquote {\bibinfo {title} {Quantum
  simulators},}\ }\href {\doibase 10.1126/science.1177838} {\bibfield
  {journal} {\bibinfo  {journal} {Science}\ }\textbf {\bibinfo {volume}
  {326}},\ \bibinfo {pages} {108--111} (\bibinfo {year} {2009})}\BibitemShut
  {NoStop}%
\bibitem [{\citenamefont {Jordan}\ and\ \citenamefont
  {Wigner}(1928)}]{Jordan:28a}%
  \BibitemOpen
  \bibfield  {author} {\bibinfo {author} {\bibfnamefont {P.}~\bibnamefont
  {Jordan}}\ and\ \bibinfo {author} {\bibfnamefont {E.}~\bibnamefont
  {Wigner}},\ }\bibfield  {title} {\enquote {\bibinfo {title} {{\"{U}ber das
  Paulische \"{A}quivalenzverbot}},}\ }\href {\doibase 10.1007/BF01331938}
  {\bibfield  {journal} {\bibinfo  {journal} {Z. Physik}\ }\textbf {\bibinfo
  {volume} {47}},\ \bibinfo {pages} {631--651} (\bibinfo {year}
  {1928})}\BibitemShut {NoStop}%
\end{thebibliography}%

\appendix

\section{Transverse Ising Model with Boundary Term Spectrum}\label{sec:lieb}

The relevant model on $l$ qubits is 
\begin{align}
	H_l(s)=-(1-s) \sum_{i=1}^l [Z]_i - s \bigg([X]_1+ \sum_{i=1}^{l-1} [X]_i [X]_{i+1}\bigg).
\end{align}
Because of the single-qubit term on the first qubit, both the initial and final Hamiltonian are nondegenerate. It is convenient to add an extra qubit, called the $0$th qubit, and consider the Hamiltonian
\begin{align}
	H_l^\prime(s)=-(1-s) \sum_{i=1}^l [Z]_i - s \bigg(\sum_{i=0}^{l-1} [X]_i [X]_{i+1}\bigg)
\end{align}
Then, when we restrict to the $+1$ eigenspace of $[X]_0$ ($[X]_0$ commutes with the Hamiltonian) we will obtain $H_l(s)$. Also note that if we conjugate $H_l(s)$ by $\prod_{i=1}^l [Z]_i$, we will obtain $H_l(s)$ with the $[X]_1$ term flipped in sign. Therefore the $-1$ eigenvalue of the $[X]_0$ eigenspace has the exact same spectrum as the $+1$ eigenvalue of the $[X]_0$ eigenspace. Thus we know that $H_l^\prime(s)$ will have exactly the same spectrum as $H_l(s)$ but will be two-fold degenerate, with the degeneracy corresponding to the eigenvalue of $[X]_0$. Thus, if we are interested in the gap of $H_l(s)$ we can work equally well with $H_l^\prime(s)$, which we now assume.

The computation of the energy spectrum of $H_l^\prime(s)$ follows the techniques of Lieb, Schultz, and Mattis~\cite{Lieb:61a}. After a Jordan-Wigner transform~\cite{Jordan:28a}, the Hamiltonian can be written as
\begin{align}\label{eq:AB}
	H_l^\prime(s)=& -s\sum_{i=0}^{l-1}\big(c_i^\dag - c_i^{\vphantom{\dag}}\big)\big(c_{i+1}^{\dag} + c_{i+1}^{\vphantom{\dag}}\big) \nonumber \\
	& \quad -(1-s)\sum_{i=1}^l \big(2 c_i^\dag c_i^{\vphantom{\dag}} -I\big)
\end{align}
We can express this as 
\begin{align}
	H_l^\prime(s)=\sum_{i,j=0}^l \Big[A_{i,j} c_i^\dag c_j^{\vphantom{\dag}} + 
	\frac{1}{2} \big( B_{i,j}  c_i^\dag c_j^\dag + B_{j,i} c_i c_j\big)\Big]+ \Gamma I
\end{align}
where $A$ is an $(l+1)\times (l+1)$ symmetric matrix given by
\begin{align}
	A_{ij}=-s(\delta_{j,i+1} + \delta_{i,j+1})-2(1-s) \delta_{i,j} (1-\delta_{i,0}), \label{eq:A}
\end{align}
$B$ a $(l+1) \times (l+1)$ antisymmetric matrix given by
\begin{align}
	B_{ij}=-s( \delta_{j,i+1} - \delta_{i,j+1}), \label{eq:B}
\end{align}
and $\Gamma=(1-s)l$. Then we can find new fermion operators, $\eta_k$ which are linear combinations of the $c_i^{\vphantom{\dag}}$ and $c_i^\dagger$ such that
\begin{align}
	H_l^\prime(s)= \sum_{k=0}^l \omega_k \eta_k^\dagger \eta_k^{\vphantom{\dag}}+\left(\frac{1}{2} \sum_{i=0}^l A_{ii} - \frac{1}{2}  \sum_{k=0}^l \omega_k+\Gamma \right)I
\end{align}
or, after simplifying,
\begin{align}
	H_l^\prime(s)= \sum_{k=0}^l \omega_k \left(\eta_k^\dagger \eta^{\vphantom{\dag}}_k-\frac{1}{2}\right)
\end{align}
with the $\omega_k$s being the square roots of the eigenvalues of $4M$~\cite{Lieb:61a}, where
\begin{align}
	M = \frac{1}{4} (A+B)(A-B) \,.
\end{align}
Note the second Hamiltonian follows by direct calculation, but also because of the traceless nature of $H_l^\prime(s)$. Explicitly, we find that
\begin{align}
	M_{ij} = &\ s^2 \delta_{i,j}(1-\delta_{i,l}) + (1-s)^2 \delta_{i,j} (1-\delta_{i,0}) \nonumber \\
	&\quad +  s(1-s)(\delta_{i,j+1} + \delta_{j,i+1}) \,.
\end{align}

It is easy to check that the (unnormalized) vector
\begin{align}
	|\Phi\rangle = \sum_{i=0}^l \left(\frac{s-1}{s} \right)^{l-i} |i\rangle 
\end{align}
is an eigenvector of $M$ with eigenvalue $0$ for all values of $s$. This implies that there is always a mode with zero energy, which is exactly a consequence of the fact that the Hamiltonian is always two-fold degenerate. As argued above, we can ignore this fact.

We can find the eigenvalues of $M$ by using an ansatz of the form
\begin{align}
	|v\rangle = \sum_{i=0}^l 
	\sin\Bigl[ \theta \bigl(i+\tfrac{1}{2} \bigr)+\phi \Bigr] |i\rangle \,.
\end{align}
If we apply $M$ to this vector then we obtain three equations: two boundary terms and a term from the bulk. Define $\gamma = s(1-s)$. The bulk term gives us the equation
\begin{align}
	\gamma & \sin \Bigl[ \theta \bigl(i-\tfrac{1}{2} \bigr)+\phi \Bigr]
	+(1-2\gamma - \lambda)\sin \Bigl[ \theta \bigl(i+\tfrac{1}{2} \bigr)+\phi \Bigr] \nonumber \\
	&\quad + \gamma \sin \Bigl[ \theta \bigl(i+\tfrac{3}{2} \bigr)+\phi \Bigr] = 0 \,,
\end{align}
where $\lambda$ is the eigenvalue. After some tedious math, this can be turned into
\begin{align}
	\lambda = (1-2\gamma) + 2 \gamma \cos \theta \,.
\end{align}
The two boundary terms give the equations
\begin{align}
	s^2 \sin \bigl( \tfrac{\theta}{2}+\phi \bigr) + 
	\gamma \sin \bigl( \tfrac{3\theta}{2} + \phi \bigr) 
	= \lambda \sin \bigl( \tfrac{\theta}{2}+\phi \bigr) 
\end{align}
and
\begin{align}
	\gamma & \sin \Bigl[ \theta \bigl(l-\tfrac{1}{2}\bigr)+\phi \Bigr]
	+ (1-s)^2 \sin \Bigl[ \theta \bigl(l+\tfrac{1}{2}\bigr)+\phi \Bigr]  \nonumber \\
	&\quad = \lambda \sin \Bigl[ \theta \bigl(l+\tfrac{1}{2}\bigr)+\phi \Bigr] \,.
\end{align}
This first equation can be rearranged to yield
\begin{align}
	(1-s) \sin \bigl(\tfrac{\theta}{2}+\phi \bigr) = s \sin \bigl( \tfrac{\theta}{2}-\phi \bigr) \label{eq:firstbound}
\end{align}
while the second one can be manipulated to become
\begin{align}
	(1-s) \sin \Big[ \theta \bigl(l+\tfrac{3}{2} \bigr) +\phi \Bigr]
	= -s \sin \Bigl[ \theta \bigl(l+\tfrac{1}{2} \bigr) +\phi \Bigr].
\end{align}
Solving these equations for $\frac{1-s}{s}$ we obtain the equation
\begin{align}
	\frac{\sin \bigl( \frac{\theta}{2} -\phi \bigr)}{\sin \bigl( \frac{\theta}{2}+\phi \bigr)} 
	= \frac{-\sin \Bigl[ \theta \bigl(l+\frac{1}{2} \bigr) +\phi \Bigr]}
	{\sin \Bigl[ \theta \bigl(l+\frac{3}{2} \bigr) +\phi \Bigr]}.
\end{align}
This can be reduced to
\begin{align}
	\cos \bigl[\theta(l+2)\bigr] = \cos (\theta l),
\end{align}
which has solutions for
\begin{align}
	\theta=\frac{\pi k}{l+1}
\end{align}
where $k$ is an integer. When $k=0$, this equation does not have a solution for general $s$. Fortunately, we already have the $k=0$ eigenvector, so we have found the relevant nonzero eigenvalues.
For $k \neq 0$, the phase shift $\phi$ can be found by solving the transcendental equation of Eq.~(\ref{eq:firstbound}). 

Thus we see that the eigenvalues are
\begin{align}\label{eq:energy}
	\omega_k(s)^2 = 4\Bigl(1-2s(1-s)
	\Bigl[1-\cos\bigl( \tfrac{k \pi}{l+1}\bigr)\Bigr] \Bigr)
\end{align}
for $k=1,2,\dots,l$. 
Each of these is minimized for $s=1/2$, for which the eigenvalues become
\begin{align}
	\omega_k(1/2)^2 = 2+2\cos\bigl( \tfrac{k \pi}{l+1}\bigr) 
	= 4 \cos^2 \bigl( \tfrac{k \pi}{2(l+1)} \bigr) \,.
\end{align}
From this we see that $\omega_k$'s smallest value is $O(1/l)$, occurring when $k=l$.

\end{document}